\documentclass[twocolumn,a4paper]{revtex4-1}
\usepackage{hyperref}
\usepackage{lipsum}
\usepackage{graphics}
\usepackage{graphicx}
\usepackage{amsmath}
\usepackage{xcolor}
\usepackage{float}
\newcommand{\be}{\begin{equation}}
\newcommand{\ee}{\end{equation}}
\newcommand{\ba}{\begin{eqnarray}}
\newcommand{\ea}{\end{eqnarray}}

\begin{document}
\title{Dissociation and thermodynamical properties of heavy quarkonia in an anisotropic strongly coupled hot QGP: using baryonic chemical potential}
 
\author{Siddhartha Solanki$^{a}$}
\author{Manohar Lal$^{a}$} 
\author{Rishabh Sharma}
\author{Vineet Kumar Agotiya$^{a}$}
\email{agotiya81@gmail.com}
\affiliation{$^a$Department of Physics, Central University of Jharkhand, Ranchi, India, 835-222}
%\date{\today}

\begin{abstract}
We extended the recent work Phys. Rev. D 97(9), 094033 (2018) to investigate  quarkonium dissociation in presence of baryonic chemical potential ($\mu_{b}$) and anisotropy ($\xi$) using quasi-particle approach in hot quantum chromodynamics (QCD) medium. We have determined binding energy and thermal width of S-states of charmonia and bottomonia for $n$=1 and $n$=2 (radial quantum number) with anisotropic parameter ($\xi$) and baryonic chemical potential. We have also determined the effects of baryonic chemical potential and anisotropy on mass spectra of 1S-states of quarkonia and the results obtained were consistent with theoretical and experimental works. But the key result obtained was dissociation temperature of the S-states with the effect of $\mu_{b}$ and $\xi$. At last, we have calculated the thermodynamical properties of QGP (i.e., pressure, energy density and speed of sound) using the parameter $\xi$ and $\mu_{b}$, which is the main key to study suppression of the quarkonium with latest determined value of energy density $\sqrt{s_{NN}}$ after incorporating the effect of $\xi$ and $\mu_{b}$. 
\end{abstract}
\maketitle

\section{Introduction}
World's largest giant accelerator, Relativistic Heavy-Ion Collision (RHIC) situated at Brookhaven national laboratory (BNL) USA and Large Hadron Collider(LHC) at CERN Switzerland inferred that quark-gluon plasma (QGP) behave like a perfect fluid instead of non-interacting gas of quasi-parton and quasi-gluons due to the collective nature of the QGP~\cite{1,2,3}. Several signature of the QGP have been identified so far but suppression of quark-antiquark pair is one of the most important or confirming signal of the QGP formation during non-central collision of heavy-ions~\cite{4,5}. Matsui and Satz~\cite{6} was first to study dissociation of the quarkonia particularly that of charmonia (J/$\psi$) by involving the color screening in deconfined state. Both experimental and theoretical studies are going on to explore properties of the QGP and few essential refinement in the QGP study have been observed during last few decades~\cite{7,8,9}. It is well known that the quarkonium bound together by static gluon and acts as independent degree of freedom~\cite{10,11,12,13}. Light hadrons were emitted during the transition of quarkonium from one state to another state while passing through the QGP medium~\cite{14}. Various authors~\cite{14,15,16} studied the features of QCD, strong theory of interaction, at high temperature scale. Studies like~\cite{17,18,19}, were dedicated to production of the quarkonium in color evaporation model or color singlet model. The suppression of QGP through coalescence or the recombination of the partons can be found in~\cite{20,21}. Due to small velocity or large mass of the heavy quark compare to QCD scale parameters, we preferably used non-relativistic approach to study the QGP properties~\cite{22,23,24}.\\
In the non-relativistic approach, we employed non-relativistic potential which possesses both fundamental features i.e., asymptotic freedom and color confinement of the QCD. In the~\cite{25,26,27,28,29,30,31}, properties (including dissociation temperature), production and suppression (both theoretical or experimental) has been already discussed in details. In~\cite{32}, dissociation temperature of states has been investigated by using quasi-particle approach in presence of the momentum anisotropy collision. In the current work, we consider the effect of momentum space anisotropy due to the fact that during the non-central heavy ion collisions, the QGP does not possess the spatial isotropy. Also there are several other studies such as~\cite{29,30,31,33,34,35} include the anisotropic effect to explore the QGP. But the key idea in the present work is to include the effect of baryonic chemical potential along with the anisotropic one in the hot QGP medium using effective fugacity quasi-particle model. The effect of the momentum space has been incorporated through the distribution function details of can be found in~\cite{10,36,37}.\\
Further, the gluon propagator and hence dielectric permittivity were modified in the presence of anisotropy ($\xi$). The effect of chemical potential has been introduced through quasi-particle Debye mass~\cite{38,39}. In this work, we modified potential accordingly. From real part of the potential, so formed, we obtained the binding energies of charmonium and bottomonium at different values of anisotropy~\cite{12,40,41,42,43,44}. The thermal width of QGP have been derived from the imaginary part of the potential~\cite{12,40,41,42,43,44}. In studies like~\cite{45,46,47,48}, authors have calculated the dissociation temperature by using the criteria of thermal width. This idea enlightened us to study binding energy and thermal width of QGP particularly at high baryon density (baryonic chemical potential). The effect of baryonic chemical potential and anisotropy significantly revise values of dissociation temperature. The thermodynamical behavior of the QGP have also been studied in the presence of $\mu_{b}$ and $\xi$. Various thermodynamical quantities of QGP such as pressure, energy density and speed of sound have been studied. These quantities played a vital role to study the Suppression of the QGP which is regarded as the most prominent signal for the existence of the QGP.\\
The present manuscript is organized in the following manner. A brief discussion about the screening between the quark-antiquark pair (Debye Screening) in the presence of temperature and $\mu_{b}$ has been provided in the section-II. In section-III, we explain about the quark-antiquark potential and its modification through Fourier Transform. We briefly explained about the inclusion of momentum space anisotropy in the medium modified form of Cornell potential in Section-IV. The binding energies of various quarkonium states have been discussed in Section-V. In section-VI, the dissociation criteria have been briefly explained. Thermal width of the quarkonia has been obtained and discussed in section-VII. Mass spectra of the charmonium and bottomonium have been calculated in section-VIII. The effect of $\mu_{b}$ and $\xi$ on the nature of thermodynamical properties ($C^{2}_{s}$, $\epsilon_{s}$ and P) has been discussed in section-IX. Finally we have concluded our work in section-X.
%%%%%%%%%%%%%%%%%%%%%%%%%%%%%%%%%%%%%%%%%%%%%%%%%%%%%%%%%%%%%%%%%%%%%%%%%%%%%%%%%%%%%%
\begin{figure*}
    \vspace{2cm}   
    \includegraphics[height=6.5cm,width=6.5cm]{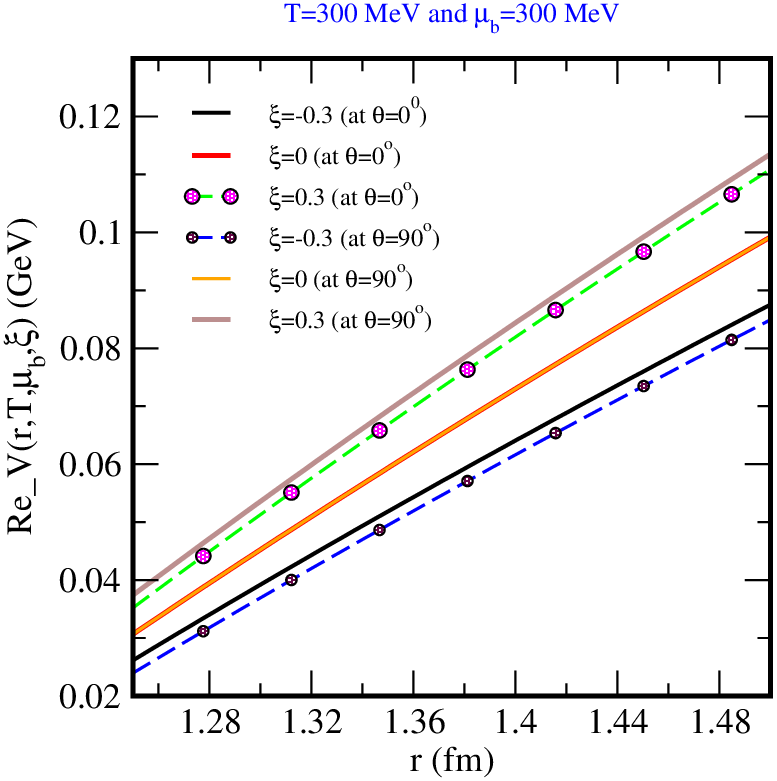}
    \hspace{2cm}
    \includegraphics[height=6.5cm,width=6.5cm]{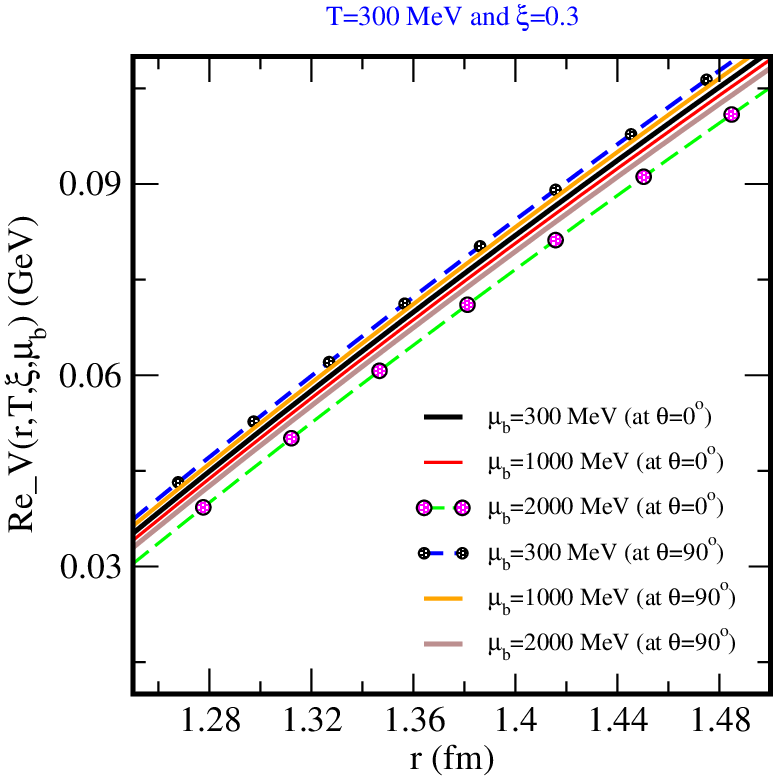}
\caption{Variation of real potential with distance (r in Fermi) at different values of anisotropy (left panel) and at different values of baryonic chemical potential (right panel) in both parallel and perpendicular case.}
\label{Fig.1}
 \vspace{3cm}  
\end{figure*}
%%%%%%%%%%%%%%%%%%%%%%%%%%%%%%%%%%%%%%%%%%%%%%%%%%%%%%%%%%%%%%%%%%%%%%%%%%%%%%%%%%%%%%%%%%
%%%%%%%%%%%%%%%%%%%%%%%%%%%%%%%%%%%%%%%%%%%%%%%%%%%%%%%%%%%%%%%%%%%%%%%%%%%%%%%%%%%%%%
\begin{figure*}
    \vspace{2cm}   
    \includegraphics[height=6.9cm,width=6.5cm]{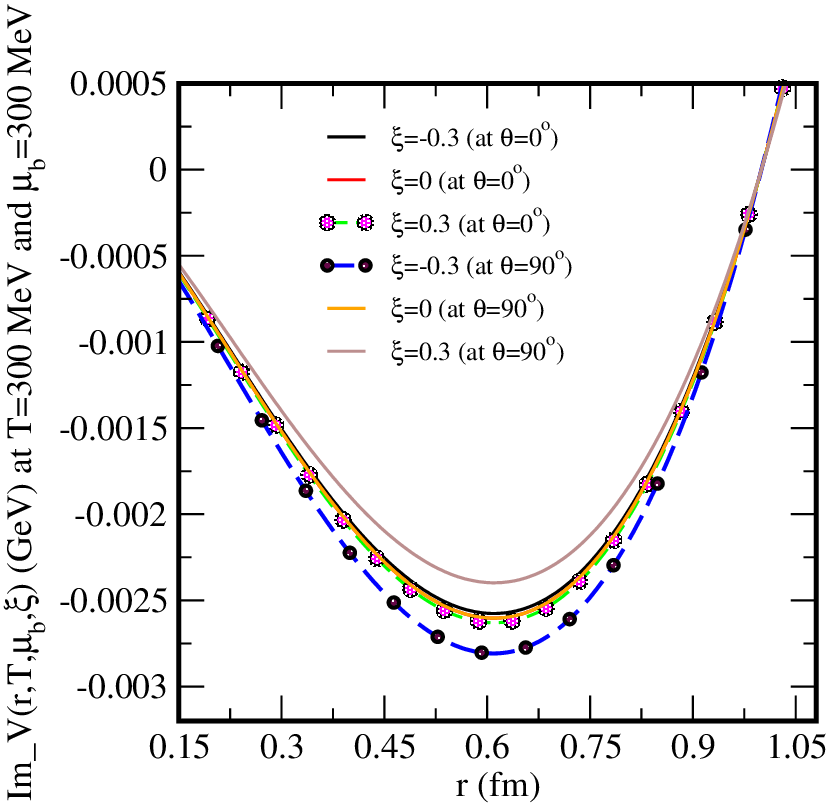}
    \hspace{2cm}
    \includegraphics[height=6.5cm,width=6.5cm]{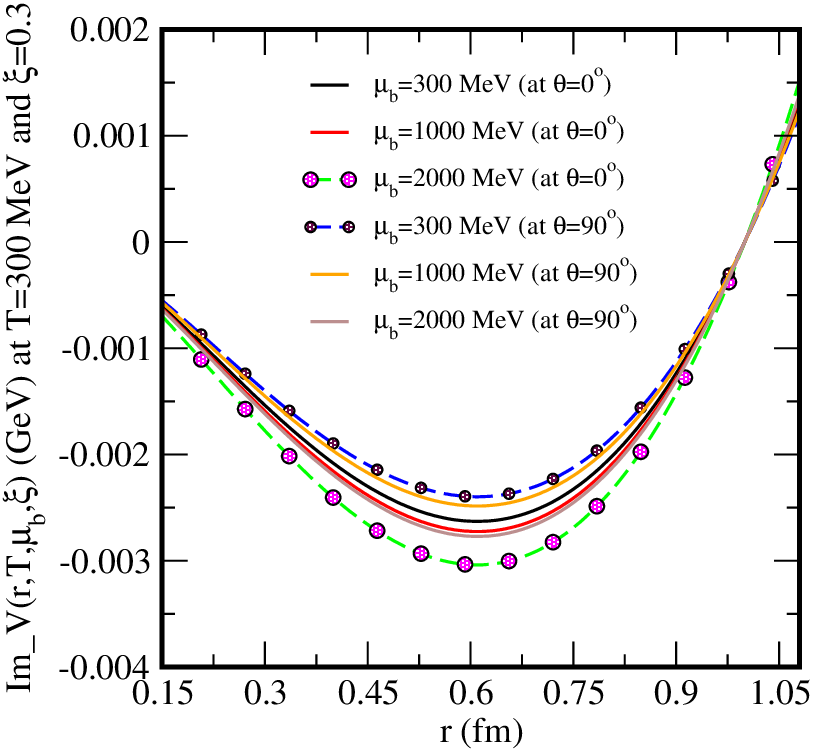}
\caption{Variation of imaginary potential with distance (r in Fermi) at different values of anisotropy (left panel) and at different values of baryonic chemical potential (right panel) in both parallel and perpendicular case.}
\label{Fig.2}
 \vspace{3cm}  
\end{figure*}
%%%%%%%%%%%%%%%%%%%%%%%%%%%%%%%%%%%%%%%%%%%%%%%%%%%%%%%%%%%%%%%%%%%%%%%%%%%%%%%%%%%%%%%%%%

\section{Study of quasi-particle Debye mass with baryonic chemical potential and temperature}
Unlike quantum electrodynamics, the Debye mass ($m_{D}$) in case of QCD is non-perturbative and gauge invariant. The leading order Debye mass in QCD coupling at high temperature has been known from the long time and perturbative in nature. Rebhan~\cite{49} has defined the Debye mass by seeing the pole of the static propagator which is relevant instead of the time-time component of the gluon self energy and obtained a Debye mass which is gauge independent. This is due to the fact that the pole of self energy does not depend on the choice of the gauge. The Debye mass was calculated for QGP, at high temperature in next to leading order (NLO) in QCD coupling from correlation of two polyakov loop by Braaten and Nieta~\cite{50}, this result agrees with the HTL result~\cite{49}. It was pointed out by Arnold and Yaffe~\cite{51} that the physics of confined magnetic charge has to be known in order to understand the contribution of O$(g^2 T)$ to the Debye mass in QCD, it was also pointed out by them that the Debye mass as a pole of gluon propagator, no longer holds true. Importantly in lattice QCD, the definition of Debye mass itself encounter difficulty due to the fact that unlike QED the electric field correlators are not gauge invariant in QCD.\\ 
The proposal of this problem is based on effective theories obtained by dimensional reduction~\cite{52}, spatial correlation 
function of gauge-invariant meson energy and the behavior of color singlet free energies~\cite{53} has been made Burnier and 
Rothkopf~\cite{54} has attempted to defined a gauge invariant mass from a complex static in medium heavy-quark potential obtained from lattice QCD. Several attempts has been made to capture all the interaction effects present in hot QCD equation of state (EoS) in terms of non-interacting quasi-partons (quasi-gluons and quasi-quarks). These quasiparton are the excitations of the interacting quarks and gluons and there are several model describing the quasi-partons such as, effective mass model~\cite{55,56}, effective mass model with polyakov loop~\cite{57}, model based on PNJL and NJL~\cite{58} and effective fugacity model~\cite{59,60}. In QCD the quasi-particle model is a phenomenological model which is widely used to describe the nonlinear behavior of QGP near phase transition point. In this model a system of interacting massless quarks and gluon can be described as an ideal gas of massive non interacting quasi particle. The mass of the quasi particle is dependent on the temperature which arises due to the interaction of gluons and quarks with surrounding medium. The quasi particle retain the quantum number of the quarks and gluons~\cite{61}. In our calculation, we used the Debye mass ($m_{D}$) for the full QCD case which was given by:
\begin{widetext}
\begin{eqnarray}
\label{eq1}
\frac{m^2_D\left(T\right)}{g^2(T) T^2}=\bigg[\bigg(\frac{N_c}{3}\times\frac{6 PolyLog[2,z_g]}{\pi^2}\bigg)+\bigg(\frac{\hat{N_f}}{6}\times\frac{-12 PolyLog[2,-z_q]}{\pi^2}\bigg)\bigg]
\end{eqnarray}
\end{widetext}
and 
\begin{eqnarray}
\label{eq2}
\hat{N_f} &=& \bigg(N_f +\frac{3}{\pi^2}\sum\frac{\mu_{b}^2}{9T^2}\bigg)
\end{eqnarray}
Here, $g(T)$ is the temperature dependent two loop running coupling constant, $N_c$=3 ($SU(3)$) and $N_f$ is the number of flavor,
the function $PolyLog[2,z]$ having form, $PolyLog[2,z]=\sum_{k=1}^{\infty} \frac{z^k}{k^2}$ and $z_g$ is the quasi-gluon effective fugacity and $z_q$ is quasi-quark effective fugacity. These distribution functions are isotropic in nature,
\begin{eqnarray}
\label{eq3}
f_{g,q}=\frac{z_{g,q}exp(-\beta p)}{\left (1\pm z_{g,q}exp(-\beta p)  \right )}
\end{eqnarray}
Where, $g$ stands for  quasi-gluons and $q$ for quasi-quarks. These fugacities should not be confused with any conservation's law (number conservation) and have merely been introduced to encode all the interaction effects at high temperature QCD. Both $z_g$ and $z_q$ have a very complicated temperature dependence and asymptotically reach to the ideal value unity~\cite{60}. The temperature dependence $z_g$ and $z_q$ fits well to the form given below,
\begin {equation}
\label{eq4}
z_{g,q}=a_{q,g}\exp\bigg(-\frac{b_{g,q}}{x^2}-\frac{c_{g,q}}{x^4}-\frac{d_{g,q}}{x^6}\bigg).
\end {equation}
(Here, $x=T/T_c$ and $a$, $b$, $c$ and $d$ are fitting parameters), for both EoS$1$ and EoS$2$. EoS$1$ is the $O(g^5)$ hot QCD~\cite{51} and EoS$2$ is the $O(g^6\ln(1/g)$ hot QCD EoS~\cite{52} in the quasi-particle description~\cite{59,60} respectively. Now, the final expressions of full QCD or quasi-particle Debye mass in terms of baryonic chemical potential and temperature can be written as:
\begin{equation}
\label{eq5}
\frac{m^2_D\left(T,\mu_{b}\right)}{T^2}=\left \{\left\{\frac{N_c}{3} Q^2_g \right\}+\left \{ \bigg[\frac{N_f}{6}+\frac{1}{2\pi^2}\bigg(\frac{\mu_{b}^2}{9T^2}\bigg)\bigg]Q^2_q\right \}\right\}
\end{equation}
Where $\mu_{b}$ is baryonic chemical potential, $Q_g$ and $Q_q$ are the effective charges given by the equations:
\begin{eqnarray}
\label{eq6}
Q^2_g&=&g^2 (T) \frac{6 PolyLog[2,z_g]}{\pi^2}\nonumber\\
Q^2_q&=&g^2 (T)  \frac{-12 PolyLog[2,-z_q]}{\pi^2}.
\end{eqnarray}
In our analysis, the temperature dependent quasi-particle Debye mass, $m_D^{QP}$ for the full QCD case with $N_f$=3 has been employed to deduce the binding energy and dissociation temperature of the quarkonia states.

\section{Modification of Cornell potential using Fourier transform (FT)}
The velocity of heavy quark mass in the bound state is small because of large quark mass (m=$m_{c,b} \geq \Lambda_{QCD}$), and the binding effects in quarkonia at the value of zero temperature can be understood in terms of non-relativistic potential models~\cite{62}. At zero temperature, the vacuum potential (Cornell potential) is given as below:
\begin{equation}
\label{eq7}
{\text V(r)} = -\frac{\alpha}{r}+\sigma r
\end{equation}
Where, $\sigma$ and $\alpha$ denotes string tension and two loop coupling constant respectively. Since the one dimensional vacuum potential defined by eq.(\ref{eq7}) is valid at zero temperature, so to study the QGP at finite temperature, the modification to the Cornell potential is required and this is done  by using Fourier transform (FT) and the medium modification enters to this heavy quark potential V(k) via FT~\cite{63} as below:
\begin{equation}
\label{eq8}
\tilde{V}(k)=\frac{\bar{\text V}(k)}{\epsilon(k)}
\end{equation}
Where, k is the Fourier conjugate of the interquark distance (r) and the dielectric permittivity ($\epsilon(k)$) is obtain by the static limit of longitudinal part of the gluon self energy~\cite{64,65}:
\begin{equation}
\label{eq9}
\epsilon(k)\equiv\left(1+\frac{m^2_D\left(T,\mu_{b}\right)}{k^2}\right)
\end{equation}
Where, $m^2_D\left(T,\mu_{b}\right)$ is the notation of quasi-particle or full QCD Debye mass with the dependency of baryonic chemical potential and temperature defined by the eq.(\ref{eq5}) in second section of the manuscript. V($\it{k}$) is the FT of the Cornell potential in eq.(\ref{eq8}), Now making the FT of eq.(\ref{eq7}) of Cornell potential is not an easy job. So, we have considered r as distribution. Then FT of Coulombic part is straight forward to compute. The FT of linear part $\sigma r exp(-\gamma r)$ is:
\begin{equation}
\label{eq10}
FT(\sigma r exp(-\gamma r))=-\frac{i\sigma}{k\sqrt{2\pi}}\left(\frac{2}{(\gamma-ik)^{3}}-\frac{2}{(\gamma+ik)^{3}}\right)
\end{equation}
At $\gamma$=0, we found the FT of $\sigma r$ is:
\begin{equation}
\label{eq11}
FT(\sigma r)=-\frac{4\sigma}{k^{4}\sqrt{2\pi}}
\end{equation}
The medium correction to the potential after applying inverse FT reads off:
\begin{equation}
\label{eq12}
V(r)= \int \frac{d^3\mathbf{k}}{(2\pi)^{3/2}}(e^{i\mathbf{k} \cdot \mathbf{r}}-1)\tilde{V}(k)
\end{equation}
FT of Cornell potential is,
\begin{equation}
\label{eq13}
{\bar{\text V}}(k)= -\sqrt\frac{2}{\pi}\bigg(\frac{\alpha}{k^2}+2\frac{ \sigma}{k^4}\bigg)
\end{equation}
Now substituting Eq.(\ref{eq9}) and (\ref{eq13}) in the Eq.(\ref{eq8}), and employing inverse FT, we got the medium modified form of potential~\cite{64,59,66} depending upon distance (r) as below:
\begin{multline}
\label{eq14}
V(r,T,\mu_{b})=\left(\frac{2\sigma}{m^2_D\left(T,\mu_{b}\right)}-\alpha\right)\frac{exp(-m_D\left(T,\mu_{b}\right)r)}{r}\\-\frac{2\sigma}{m^2_D\left(T,\mu_{b}\right)r}+\frac{2\sigma}{m_D\left(T,\mu_{b}\right)}-\alpha m_D\left(T,\mu_{b}\right)
\end{multline}
It is also noticeable that, in hot QCD medium the expression of potential is not same as the lattice parametrized heavy quark free energy in the deconfined phase (which is basically a screened Coulomb, for the exact form we refer the reader to Reference~\cite{67} for more details). As emphasized by Dixit~\cite{68} that one dimensional FT of the Cornell potential in the medium yields the similar form as used in the lattice QCD to study the quarkonium properties which assumes one-dimensional color flux tube structure. Since the flux tube structure may expand in more dimensional~\cite{67}. Therefore, it is better to consider the three dimensional form of the medium modified form of Cornell potential which has been done exactly in the present manuscript.

\section{Quark-antiquark potential in anisotropic medium using baryonic chemical potential}
The spatial anisotropy $(\xi)$, in non-central heavy ion collisions, generates at the early stages of QGP. As the system evolves with time, different pressure gradients produce in different direction which maps the spatial anisotropy to the momentum anisotropy. The anisotropy in this manuscript has been introduced at the particle phase space distribution level. Applying the method used in references~\cite{69,70,71}, the distribution function of anisotropy has been obtained from isotropic one by stretching and squeezing it in one of the direction in the momentum space as: 
\ba
f({\mathbf{p}})\rightarrow f_{\xi}({\mathbf{p}}) = C_{\xi}~f(\sqrt{{\bf p}^{2} + \xi({\bf p}\cdot{\bf \hat{n}})^{2}})
\label{eq15}
\ea
Where, $f({\mathbf{p}})$ represents isotropic distribution function as in \cite{60,72}, ${\mathbf{\hat{n}}}$ is unit vector in momentum anisotropy direction [example- In squeezing ($\xi>0$ or oblate case) and in stretching ($-1<\xi<0$ or prolate case) in the ${\mathbf{\hat{n}}}$ direction]. Whereas, $\xi$ denotes, anisotropy of the medium. Various EoSs effects enter through the Debye screening mass ($m_D$), to make Debye mass similar in both isotropic ($\xi=0$) and anisotropic ($\xi \neq 0$)~\cite{73}, we used normalization constant, $C_{\xi}$ as given below:
\ba
C_{\xi}= \left\{ \begin{array}{cl}
\frac{\sqrt{|\xi|}}{tanh^{-1}\sqrt{|\xi|}}~~~if~~~ -1 \leq \xi< 0\\
\frac{\sqrt{\xi}}{tan^{-1}\sqrt{\xi}}~~~if~~~ \xi \geq 0
\end{array} \right.
\label{eq16}
\ea
The limit of $\xi$ is small then we have:
\ba
C_{\xi}= \left\{ \begin{array}{cl}
1-{\frac{\xi}{3}}+O({\xi^{\frac{3}{2}}})~~~if~~~ -1 \leq \xi< 0\\
1+{\frac{\xi}{3}}+O({\xi^{\frac{3}{2}}})~~~if~~~ \xi \geq 0
\end{array} \right.
\label{eq17}
\ea
In the presence of dissipative anisotropic hot QCD medium, we have modified the potential after considering the assumption given in references~\cite{74,75,76}. We have already discussed it in details in section-II, how to obtain the in medium modification of the heavy quark potential with dielectric permittivity $\epsilon(k)$. We have already calculated the expression of the FT of Cornell potential (Eq.(\ref{eq13})) in section-II.
%%%%%%%%%%%%%%%%%%%%%%%%%%%%%%%%%%%%%%%%%%%%%%%%%%%%%%%%%%%%%%%%%%%%%%%%%%%%%%%%%%%%%%%%%%%%%%%%%%%%%%%%%%%%%%%%%%%%%%
\begin{figure*}
    \vspace{2cm}   
    \includegraphics[height=6.5cm,width=6.5cm]{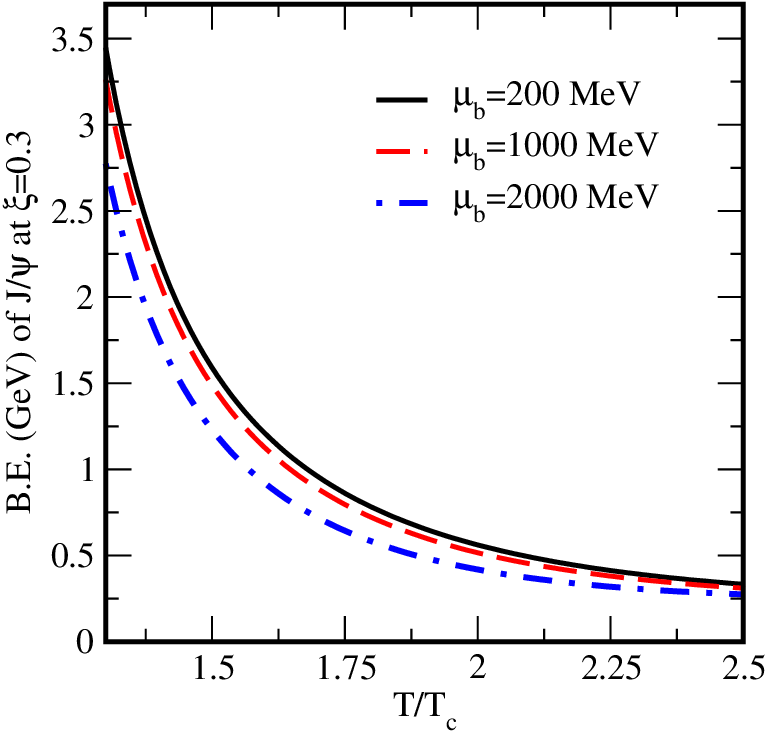}
    \hspace{2cm}
    \includegraphics[height=6.5cm,width=6.5cm]{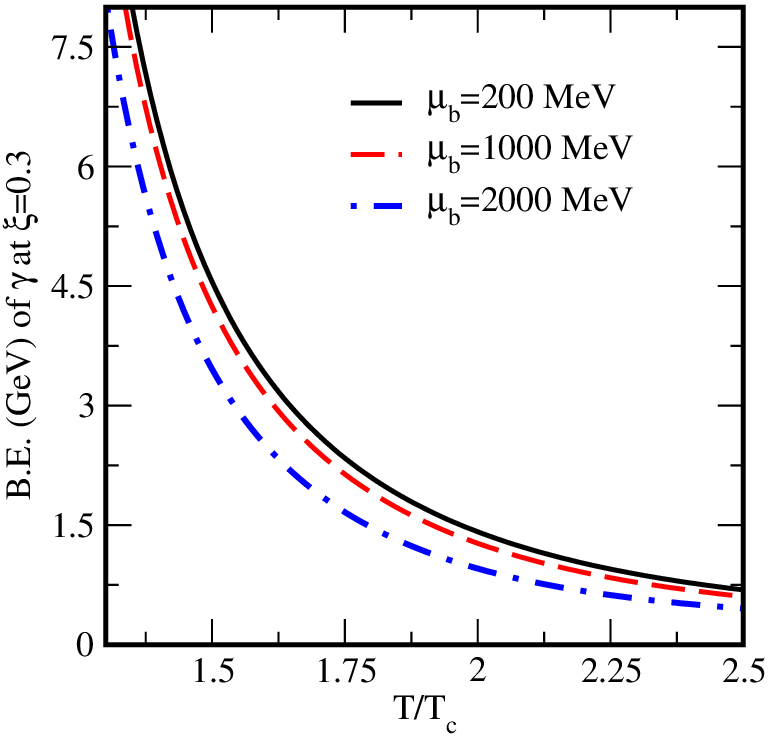}
    \vspace{2mm}
\caption{Shows the variation of binding energy of the $J/\psi$ (left panel) and $\Upsilon$ (right panel) with T/$T_{c}$ at different values of baryonic chemical potential ($\mu_{b}$) when the value of $\xi$ is fixed.}
\label{Fig.3}
 \vspace{3cm}  
\end{figure*}
%%%%%%%%%%%%%%%%%%%%%%%%%%%%%%%%%%%%%%%%%%%%%%%%%%%%%%%%%%%%%%%%%%%%%%%%%%%%%%%%%%%%%%%%%%%%%%%%%%%%%%%%%%%%%%%%%%%%%%%%%%%%
While modifying potential, the foremost thing is to calculate the dielectric permittivity $\epsilon(\bf k)$. To estimate dielectric permittivity, we have following two approaches: (I) With the help of gluon self energy in at finite temperature QCD~\cite{77,78} and (II) by the application of semi-classical transport theory~\cite{79,80,81}. By exploiting any of these two methods, one can find the gluon self energy tensor $(\Pi^{\mu\nu})$ and then the static gluon propagator represents inelastic scattering of an off-shell gluon to a thermal gluons:
\ba
\Delta^{\mu\nu}(\omega, {\bf k})= k^{2}g^{\mu\nu} - k^{\mu} k^{\nu}+ \Pi^{\mu\nu}(\omega, {\bf k}) 
\label{eq18}
\ea
Where, $\omega$ is the frequency, and the gluon self energy tensor is symmetric and transverse in nature, i.e., $\Pi^{\mu\nu}(\omega,{\bf k})=\Pi^{\nu\mu}(\omega,{\bf k})$ and follow the Ward's identity. 
\begin{multline}
\label{eq19}
\Pi^{\mu\nu}(\omega,{\bf k})=g^{2}\int\frac{d^{3}p}{(2\pi)^{3}}u^{\mu}\frac{\partial f(p)}{\partial p^{\beta}}\left[ g^{\nu\beta}-\frac{u^{\nu}k^{\beta}}{uk+i\epsilon} \right]
\end{multline}
The term, $\mu^{\mu}=(1+\frac{\bf k}{|\bf k|})$ is a light like vector define the propagation of plasma particle in space time, and in quantum chromodynamics plasma  whereas $f(p)$ is denoted the arbitrary particle distribution function. In Fourier space, gluon propagators with real and imaginary part of potential obtained from dielectric tensor of temporal component, thus, becomes:
\ba
\epsilon^{-1}({\bf k}) = -\lim_{\omega \to 0}k^2\Delta^{00}(\omega,{\bf k})
\label{eq20}
\ea
Where, $\Delta^{00}$ represents the static limit of the $00$ components of the gluon propagators in the Coulomb gauge.
%%%%%%%%%%%%%%%%%%%%%%%%%%%%%%%%%%%%%%%%%%%%%%%%%%%%%%%%%%%%%%%%%%%%%%%%%%%%%%%%%%%%%%%%%%%%%%
\begin{figure*}
    \vspace{2cm}   
    \includegraphics[height=6.5cm,width=6.5cm]{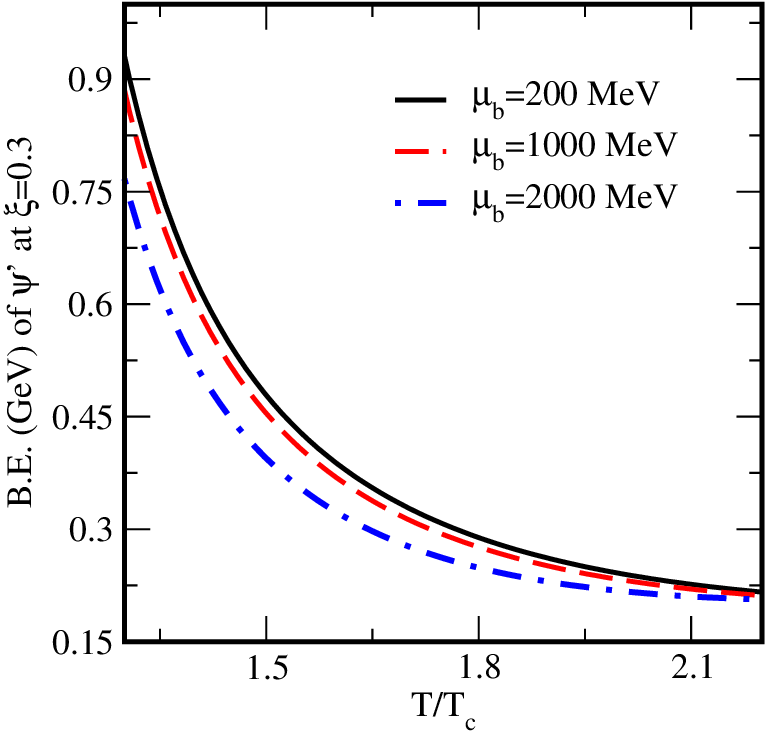}
    \hspace{2cm}
    \includegraphics[height=6.5cm,width=6.5cm]{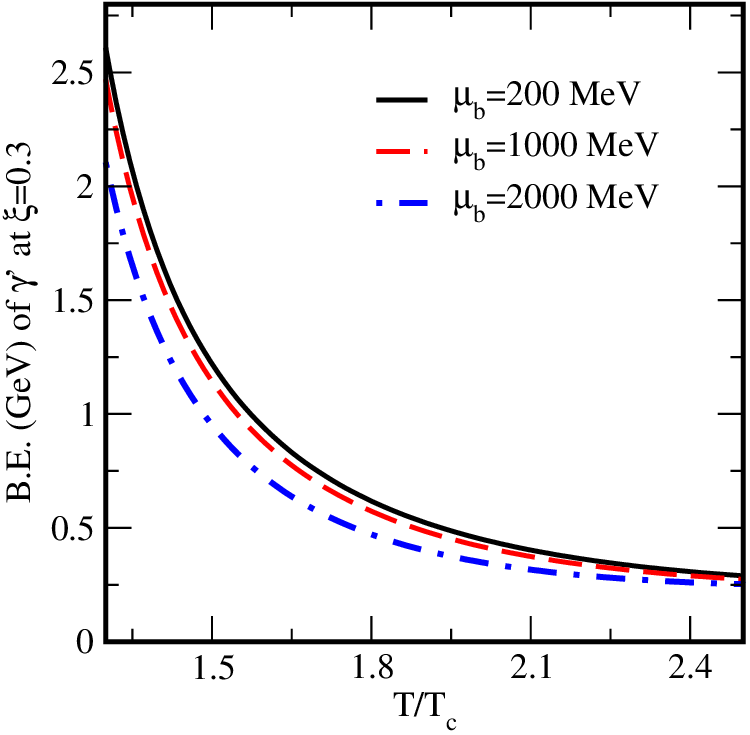}
    \vspace{2mm}
\caption{Shows the variation of binding energy of the $\psi^{\prime}$ (left panel) and $\Upsilon^{\prime}$ (right panel) with T/$T_{c}$ at different values of baryonic chemical potential ($\mu_{b}$) where the value of $\xi$ is fixed.}
\label{Fig.4}
 \vspace{3cm}  
\end{figure*}
%%%%%%%%%%%%%%%%%%%%%%%%%%%%%%%%%%%%%%%%%%%%%%%%%%%%%%%%%%%%%%%%%%%%%%%%%%%%%%%%%%%%%%%%%%%%%%%
%%%%%%%%%%%%%%%%%%%%%%%%%%%%%%%%%%%%%%%%%%%%%%%%%%%%%%%%%%%%%%%%%%%%%%%%%%%%%%%%%%%%%%%%%%%%%%
\begin{figure*}
    \vspace{2cm}   
    \includegraphics[height=6.5cm,width=6.5cm]{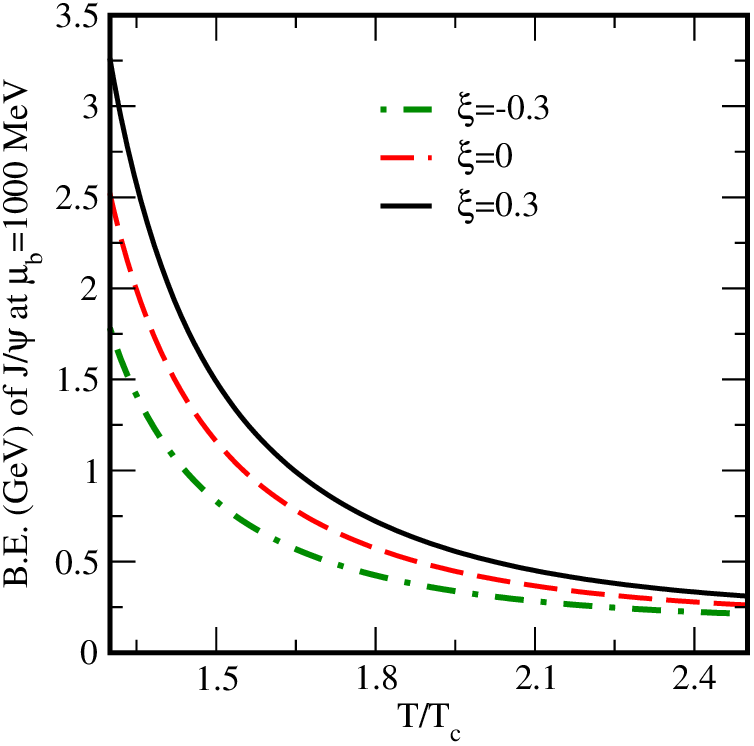}
    \hspace{2cm}
    \includegraphics[height=6.5cm,width=6.5cm]{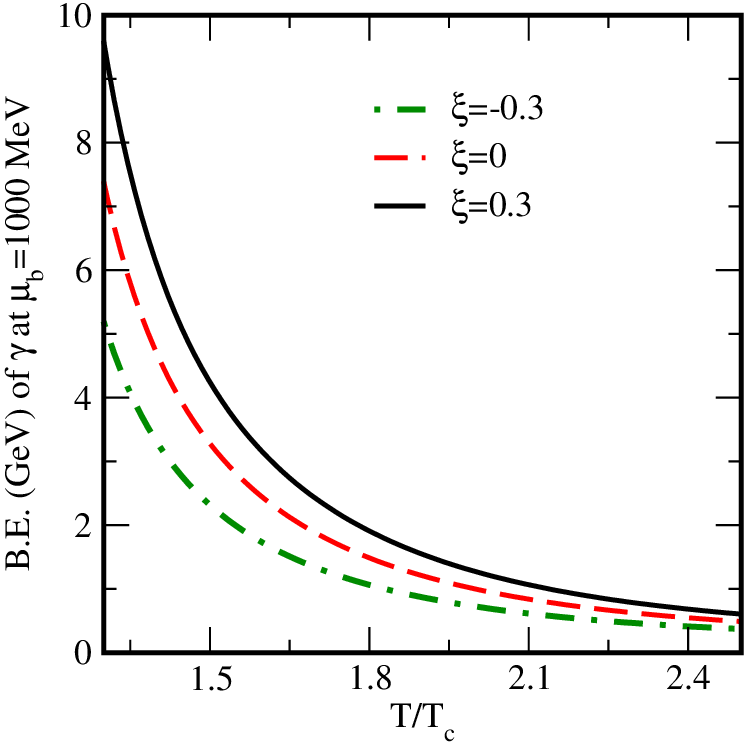}
    \vspace{2mm}
\caption{Shows the variation of binding energy of the $J/\psi$ (left panel) and $\Upsilon$ (right panel) with T/$T_{c}$ at different values of anisotropy ($\xi$) and when value of $\mu_{b}$ is fixed.}
\label{Fig.5}
 \vspace{3cm}  
\end{figure*}
%%%%%%%%%%%%%%%%%%%%%%%%%%%%%%%%%%%%%%%%%%%%%%%%%%%%%%%%%%%%%%%%%%%%%%%%%%%%%%%%%%%%%%%%%%%%%%%
%%%%%%%%%%%%%%%%%%%%%%%%%%%%%%%%%%%%%%%%%%%%%%%%%%%%%%%%%%%%%%%%%%%%%%%%%%%%%%%%%%%%%%%%%%%%%%
\begin{figure*}
    \vspace{2cm}   
    \includegraphics[height=6.5cm,width=6.5cm]{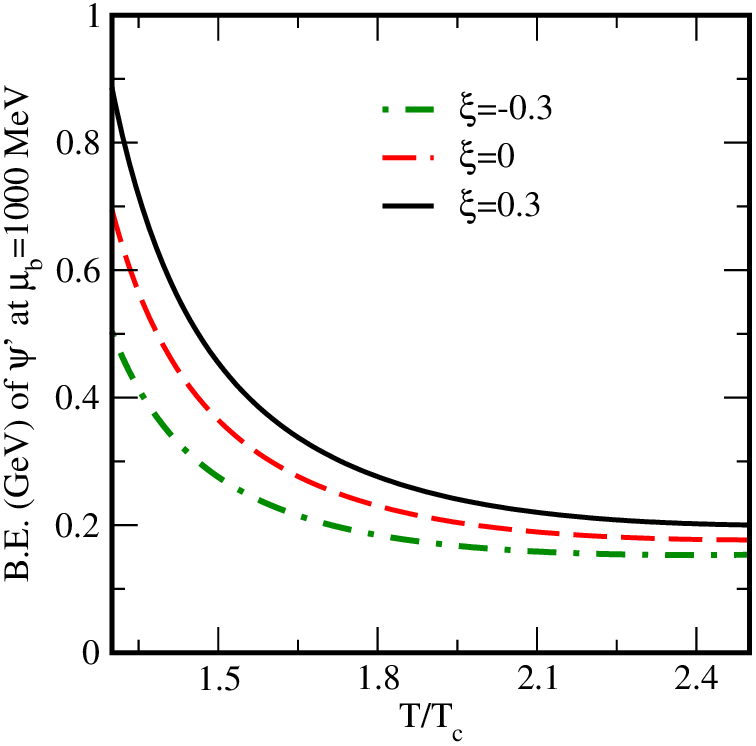}
    \hspace{2cm}
    \includegraphics[height=6.5cm,width=6.5cm]{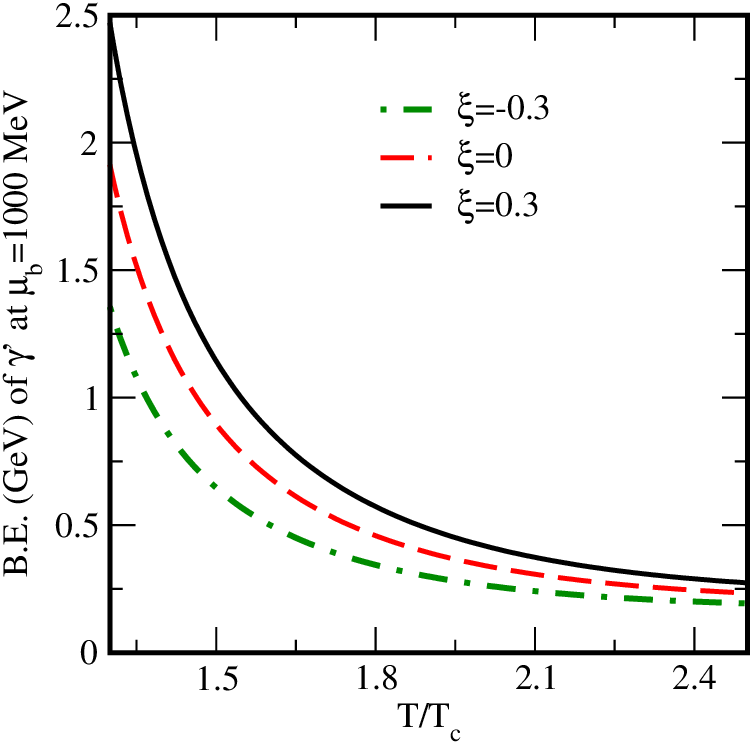}
    \vspace{2mm}
\caption{Shows the variation of binding energy of the $\psi^{\prime}$ (left panel) and $\Upsilon^{\prime}$ (right panel) with T/$T_{c}$ at different values of anisotropy ($\xi$) where the value of $\mu_{b}$ is fixed.}
\label{Fig.6}
 \vspace{3cm}  
\end{figure*}
%%%%%%%%%%%%%%%%%%%%%%%%%%%%%%%%%%%%%%%%%%%%%%%%%%%%%%%%%%%%%%%%%%%%%%%%%%%%%%%%%%%%%%%%%%%%%%%
%%%%%%%%%%%%%%%%%%%%%%%%%%%%%%%%%%%%%%%%%%%%%%%%%%%%%%%%%%%%%%%%%%%%%%%%%%%%%%%%%%%%%%%%%%%%%%
\begin{figure*}
    \vspace{2cm}   
    \includegraphics[height=6.5cm,width=6.5cm]{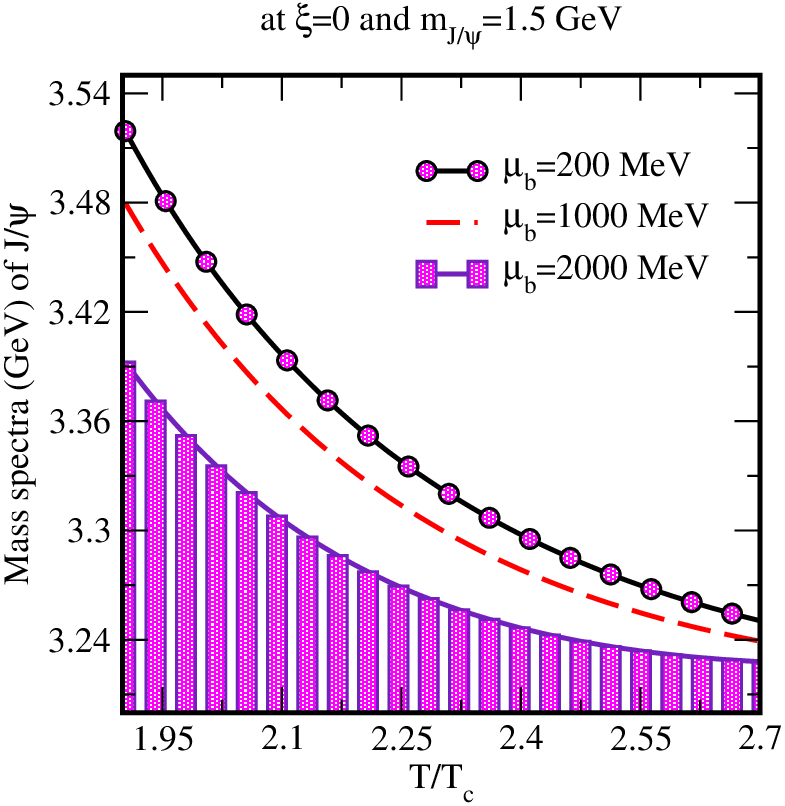}
    \hspace{2cm}
    \includegraphics[height=6.5cm,width=6.5cm]{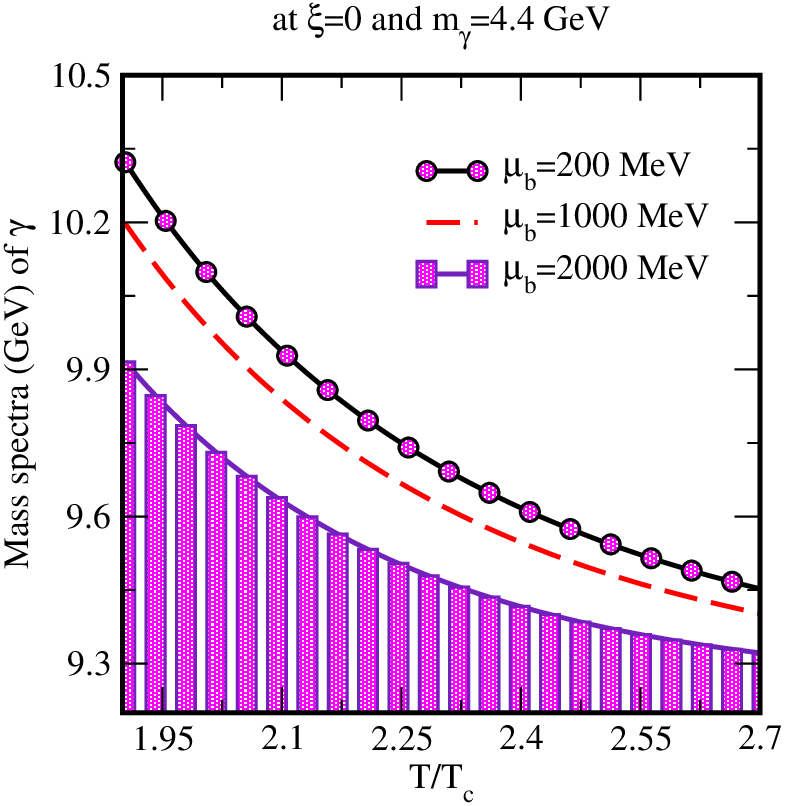}
    \vspace{2mm}
\caption{Shows the variation of mass spectra of the $J/\psi$ (left panel) and $\Upsilon$ (right panel) with T/$T_{c}$ at different values of anisotropy ($\xi$) when the value of $\mu_{b}$ is fixed.}
\label{Fig.7}
 \vspace{3cm}  
\end{figure*}
%%%%%%%%%%%%%%%%%%%%%%%%%%%%%%%%%%%%%%%%%%%%%%%%%%%%%%%%%%%%%%%%%%%%%%%%%%%%%%%%%%%%%%%%%%%%%%%%%
%%%%%%%%%%%%%%%%%%%%%%%%%%%%%%%%%%%%%%%%%%%%%%%%%%%%%%%%%%%%%%%%%%%%%%%%%%%%%%%%%%%%%%%%%%%%%%
\begin{figure*}
    \vspace{2cm}   
    \includegraphics[height=6.5cm,width=6.5cm]{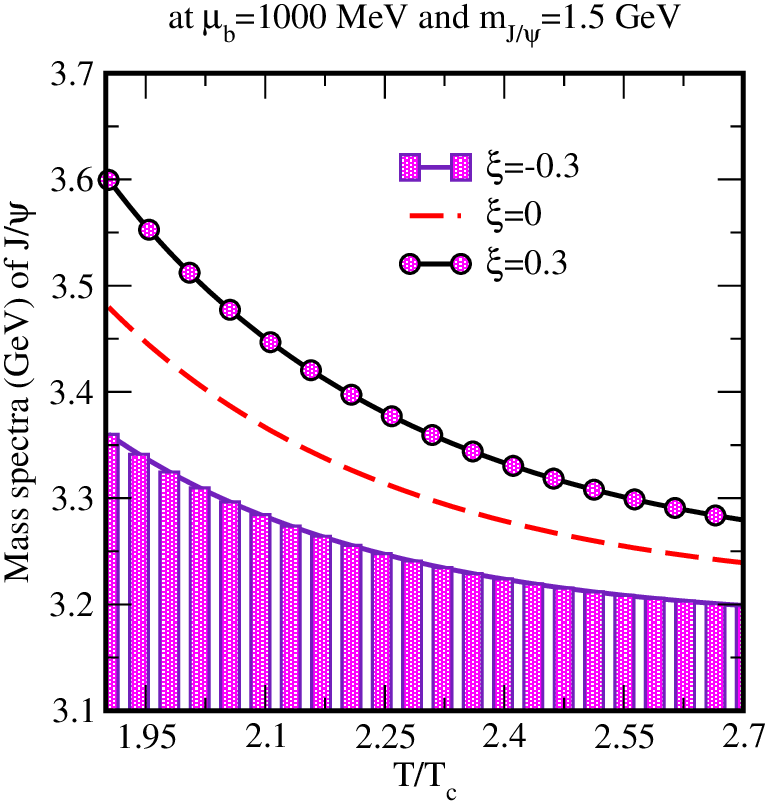}
    \hspace{2cm}
    \includegraphics[height=6.5cm,width=6.5cm]{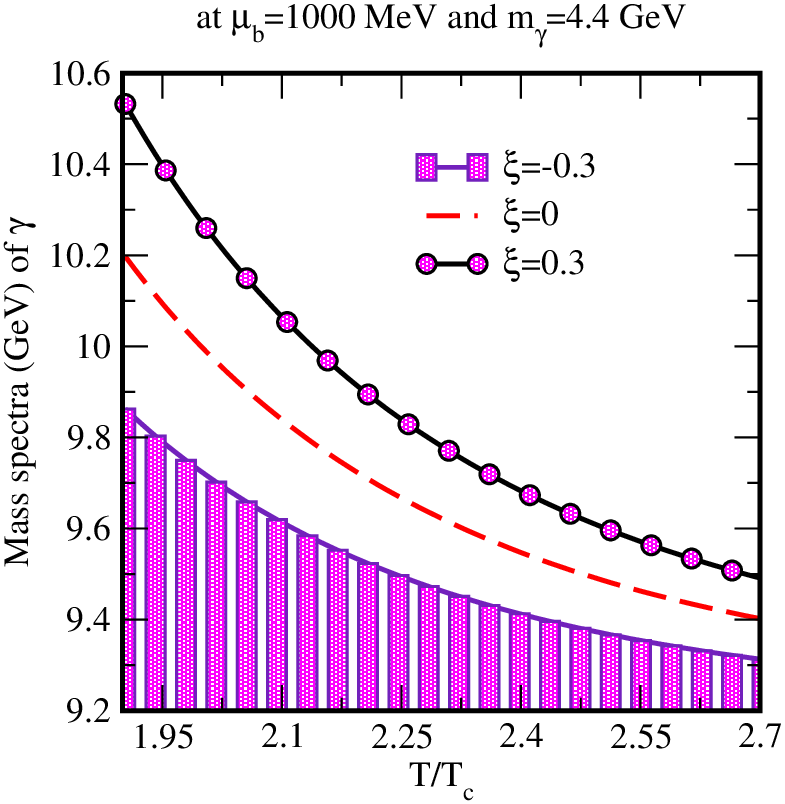}
    \vspace{2mm}
\caption{Shows the variation of mass spectra of the $J/\psi$ (left panel) and $\Upsilon$ (right panel) with T/$T_{c}$ at different values of anisotropy ($\xi$) where the value of $\mu_{b}$ is fixed.}
\label{Fig.8}
 \vspace{3cm}  
\end{figure*}
%%%%%%%%%%%%%%%%%%%%%%%%%%%%%%%%%%%%%%%%%%%%%%%%%%%%%%%%%%%%%%%%%%%%%%%%%%%%%%%%%%%%%%%%%%%%%%%%%
After performing calculation (shown in Appendix), we calculate the real and imaginary part of the temporal component of the propagator in the static limit using quasi-particle Debye mass. The temporal component of the real part of retarded propagator in the Fourier space which is required to obtain the real part of the potential in static limit ~\citep{46}, is given below as:
\begin{multline}
\label{eq21}
Re[\Delta^{00}_{R(A)}](\omega=0,{\bf k})=-\frac{1}{k^{2}+m^{2}_{D}\left(T,\mu_{b}\right)}\\-\xi\left\{\frac{1}{3(k^{2}+m^{2}_{D}\left(T,\mu_{b}\right))}-\frac{m^{2}_{D}\left(T,\mu_{b}\right)(3cos2\theta_{n}-1)}{6(k^{2}+m^{2}_{D}\left(T,\mu_{b}\right))^{2}}  \right\}
\end{multline}
Similarily, the imaginary part can be derived from the imaginary part of the temporal component of the symmetric propagator~\citep{46}, in the static limit, which is given as below:
\begin{widetext}  
\begin{multline}
\label{eq22}
Im[\Delta^{00}_{S}](\omega=0,{\bf k})+\frac{\pi T m^{2}_{D}\left(T,\mu_{b}\right)}{k(k^{2}+m^{2}_{D}\left(T,\mu_{b}\right))^{2}}=\pi T m^{2}_{D}\left(T,\mu_{b}\right)\xi\left[\frac{-1}{3k(k^{2}+m^{2}_{D}\left(T,\mu_{b}\right))^{2}}+\frac{3sin^{2}\theta_{n}}{4k(k^{2}+m^{2}_{D}\left(T,\mu_{b}\right))^{2}} \right]\\-\pi T m^{2}_{D}\left(T,\mu_{b}\right)\xi\left[\frac{2m^{2}_{D}\left(T,\mu_{b}\right)(3sin^{2}\theta_{n}-1)}{3k(k^{2}+m^{2}_{D}\left(T,\mu_{b}\right))^{3}} \right]
\end{multline}
\end{widetext}
Where,
\begin{multline}
\label{eq23}
\cos(\theta_n)=\cos(\theta_r)\cos(\theta_{pr})+\sin(\theta_r)\sin(\theta_{pr})\cos (\phi_{pr})
\end{multline}
In the above expression, $\theta_{n}$ represents angle between the particle momentum ($\bf{p}$) and the direction of anisotropy, $\theta_{r}$ denotes angle between ${\bf r}$, and ${\bf n}$. $\phi_{pr}$ and $\theta_{pr}$ are azimuthal and polar angle. Next to modify the real part of the potential, $\epsilon({\bf k})$ can be obtained using eq.(\ref{eq21}) in eq.(\ref{eq20}) as:
\begin{multline}
\label{eq24}
\epsilon^{-1}({\bf k})=\frac{k^{2}}{k^{2}+m^{2}_{D}\left(T,\mu_{b}\right)}+k^{2}\xi\\\left\{\frac{1}{3(k^{2}+m^{2}_{D}\left(T,\mu_{b}\right))}-\frac{m^{2}_{D}\left(T,\mu_{b}\right)(3cos2\theta_{n}-1)}{6(k^{2}+m^{2}_{D}\left(T,\mu_{b}\right))^{2}}\right\}
\end{multline}
Similarly for imaginary part, $\epsilon({\bf k})$ can be obtained by employing eq.(\ref{eq22}) in eq.(\ref{eq20}) as:
\begin{multline}
\label{eq25}
\frac{\epsilon^{-1}({\bf k})}{\pi~ T~ m_{D}^{2}\left(T,\mu_{b}\right)}=\bigg(\frac{k^2}{k(k^2+m_{D}^{2}\left(T,\mu_{b}\right))^2}\bigg)\\-\xi
k^2\bigg(\frac{-1}{3k(k^2+m_{D}^{2}\left(T,\mu_{b}\right))^2} \\ +\frac{3\sin^{2}{\theta_n}}{4k(k^2+m_{D}^{2}\left(T,\mu_{b}\right))^2} -\frac{2m_{D}^{2}\left(T,\mu_{b}\right)\big(3\sin^{2}({\theta_n})-1\big)}{3k(k^2+m_{D}^{2}\left(T,\mu_{b}\right))^3}\bigg)
\end{multline}
Real and imaginary part of inter-quark potential can be obtained in static limit using $\epsilon^{-1}({\bf k})$,~\cite{72}. Using eq.(\ref{eq24}) in eq.(\ref{eq12}), we can write the real part of the potential:
\begin{widetext}
\begin{multline}
Re[V(r,\theta_{r},\xi,T,\mu_{b})]=\int \frac{d^3\mathbf{k}}{(2\pi)^{3/2}}(e^{i\mathbf{k} \cdot \mathbf{r}}-1)\bigg(-\sqrt{\frac{2}{\pi}}\frac{\alpha}{k^2}-\frac{4\sigma}{\sqrt{2\pi}k^4}\bigg)\left(\frac{k^2}{k^2+m_{D}^{2}\left(T,\mu_{b}\right)}\right)\\+k^2\xi\left(\frac{1}{3(k^2+m_{D}^{2}\left(T,\mu_{b}\right))}-\frac{m_{D}^{2}\left(T,\mu_{b}\right)(3\cos{2\theta_n}-1)}{6(k^2+m_{D}^{2}\left(T,\mu_{b}\right))^2}\right)\bigg(-\sqrt{\frac{2}{\pi}}\frac{\alpha}{k^2}-\frac{4\sigma}{\sqrt{2\pi}k^4}\bigg)
\label{eq26}
\end{multline}
\end{widetext}
Where s=r$m_{D}\left(T,\mu_{b}\right)$, and after considering the limit $s\ll1$, solution to the above integral yields: 
\begin{widetext}
\begin{equation}
Re[V(r,\theta_{r},\xi,T,\mu_{b})]=\frac{s\sigma}{m_{D}\left(T,\mu_{b}\right)}\left(1+\frac{\xi }{3}\right)-\frac{\alpha m_{D}\left(T,\mu_{b}\right)}{s}\left[ 1+\frac{s^{2}}{2}+\xi\left( \frac{1}{3}+\frac{s^{2}}{16}\left( \frac{1}{3}+cos(2\theta_{r})\right)\right)\right]
\label{eq27}
\end{equation}
\end{widetext}
The imaginary potential, using eq.(\ref{eq25}) in eq.(\ref{eq12}) will be:
\begin{widetext}
\begin{multline}
Im[V(r,\theta_{r},\xi,T,\mu_{b})]=\pi T m_{D}^{2}\left(T,\mu_{b}\right)\int \frac{d^3\mathbf{k}}{(2\pi)^{3/2}}(e^{i\mathbf{k} \cdot \mathbf{r}}-1)\bigg(-\sqrt{\frac{2}{\pi}}\frac{\alpha}{k^2}-\frac{4\sigma}{\sqrt{2\pi}k^4}\bigg)\left( \frac{k}{(k^{2}+m^{2}_{D}\left(T,\mu_{b}\right))^{2}} \right)\\-\pi T m_{D}^{2}\left(T,\mu_{b}\right)\xi\int \frac{d^3\mathbf{k}}{(2\pi)^{3/2}}(e^{i\mathbf{k} \cdot \mathbf{r}}-1)\\\bigg(-\sqrt{\frac{2}{\pi}}\frac{\alpha}{k^2}-\frac{4\sigma}{\sqrt{2\pi}k^4}\bigg)\left(\frac{-k}{3(k^{2}+m^{2}_{D}\left(T,\mu_{b}\right))^{2}}+\frac{3ksin^{2}\theta_{n}}{4(k^{2}+m^{2}_{D}\left(T,\mu_{b}\right))^{2}}-\frac{2m^{2}_{D}\left(T,\mu_{b}\right)k(3sin^{2}\theta_{n}-1)}{(k^{2}+m^{2}_{D}\left(T,\mu_{b}\right))^{2}}\right)
\label{eq28}
\end{multline}
\end{widetext}
Upto leading logarithmic order, imaginary potential is:
\begin{multline}
Im[V(r,\theta_{r},T,\mu_{b},\xi)]=\frac{\alpha s^{2}T}{3}\left\{ \frac{\xi}{60}(7-9cos2\theta_{r})-1 \right\}\\log\frac{1}{s}+\frac{s^{4}\sigma T}{m^{2}_{D}\left(T,\mu_{b}\right)}\left\{\frac{\xi}{35}\left(\frac{1}{9}-\frac{1}{4}cos2\theta_{r}\right)-\frac{1}{30}\right\}\\log\frac{1}{s}
\label{eq29}
\end{multline}
Fig.\ref{Fig.1} and \ref{Fig.2} shows variation of the real and imaginary potential with distance (r) at constant temperature T=300 MeV. Left panel of the fig.\ref{Fig.1} and \ref{Fig.2}, shows the real and imaginary part of the potential for different anisotropic case: prolate $\xi$=-0.3, oblate $\xi$=0.3 and isotropic $\xi$=0 at $\mu_{b}$=300 MeV and $\theta$=$0^{o}$ (parallel case) and $\theta$=$90^{o}$ (perpendicular case). It was observed that there was an increase in real potential as one goes from prolate to oblate case. The imaginary potential decreases from prolate to oblate case in parallel case but increases for perpendicular case. The right panel of fig.\ref{Fig.1} and \ref{Fig.2} represents same variation for real and imaginary potential at constant $\xi$=0.3 for different $\mu_{b}$=300, 1000 and 2000 MeV. It was observed that the real potential increases with $r$ for different baryonic chemical potential, and imaginary potential shows decreasing pattern. In concise, potential (real and imaginary) have higher values for perpendicular case (in right panel). This indicates that anisotropy and baryonic chemical potential has significant effect on complexed valued potential.

%%%%%%%%%%%%%%%%%%%%%%%%%%%%%%%%%%%%%%%%%%%%%%%%%%%%%%%%%%%%%%%%%%%%%%%%%%%%%%%%%%%%%%%%%%%%%%%%%%5555
\begin{figure*}
    \vspace{2cm}   
    \includegraphics[height=6.5cm,width=6.5cm]{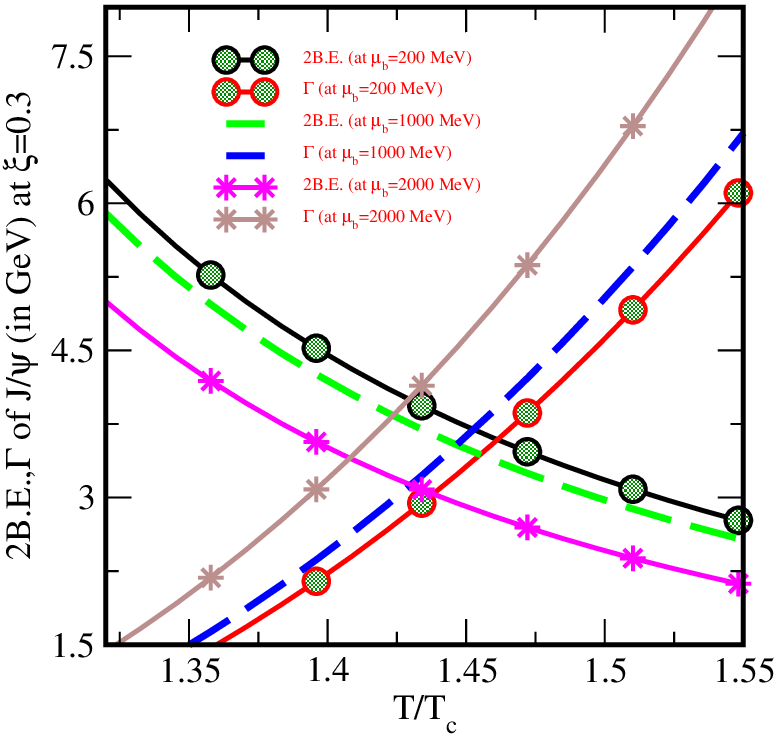}
    \hspace{2cm}
    \includegraphics[height=6.5cm,width=6.5cm]{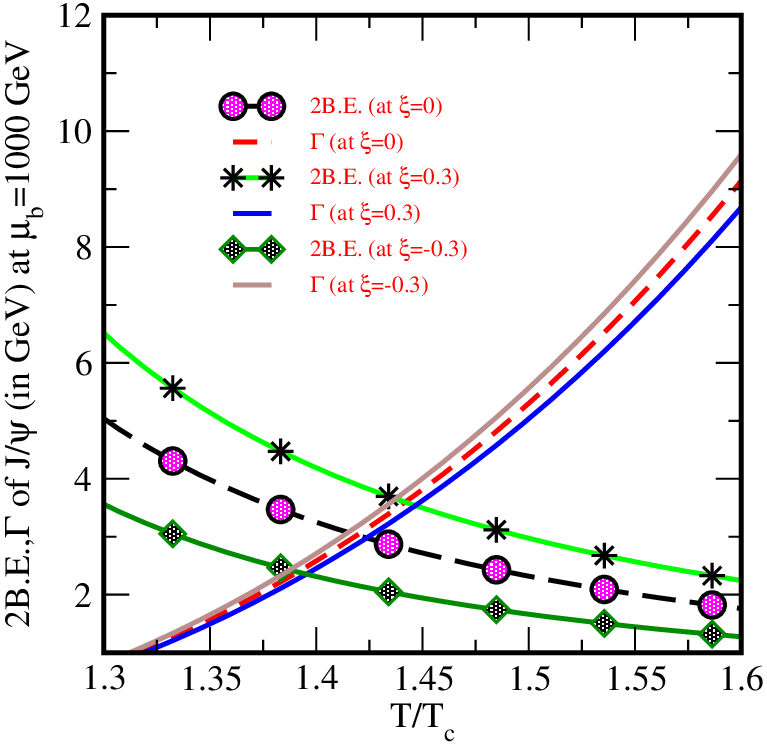}
    \vspace{2mm}
\caption{Shows the variation of 2B.E., $\Gamma$ of $J/\psi$ with $T/T_{c}$ at different values of $\mu_{b}$ (left panel) and at different values of $\xi$ (right panel).}
\label{Fig.9}
 \vspace{3cm}  
\end{figure*}
%%%%%%%%%%%%%%%%%%%%%%%%%%%%%%%%%%%%%%%%%%%%%%%%%%%%%%%%%%%%%%%%%%%%%%%%%%%%%%%%%%%%%%%%%%%%%%%%%%%%%%%%%%
%%%%%%%%%%%%%%%%%%%%%%%%%%%%%%%%%%%%%%%%%%%%%%%%%%%%%%%%%%%%%%%%%%%%%%%%%%%%%%%%%%%%%%%%%%%%%%%%%%%%
\begin{figure*}
    \vspace{2cm}   
    \includegraphics[height=6.5cm,width=6.5cm]{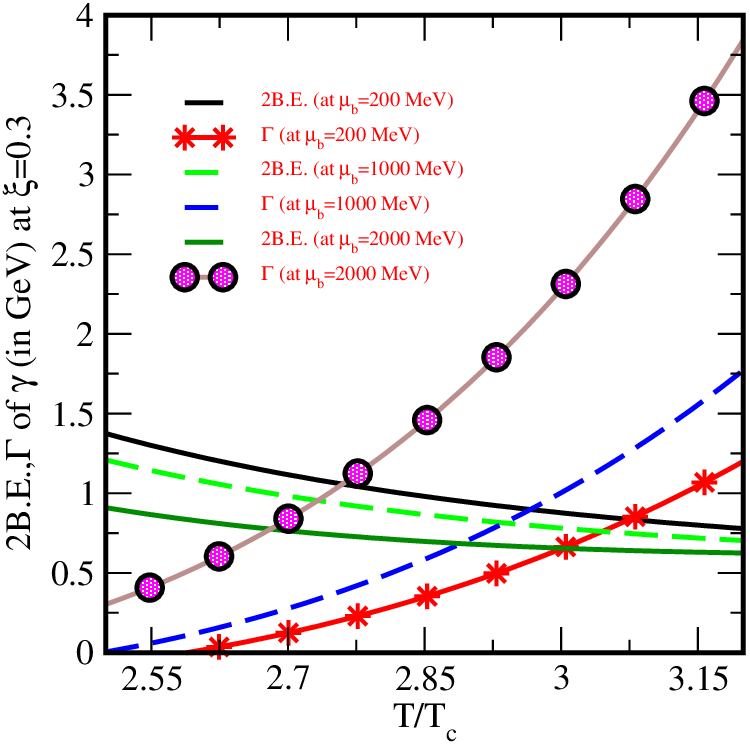}
    \hspace{2cm}
    \includegraphics[height=6.5cm,width=6.5cm]{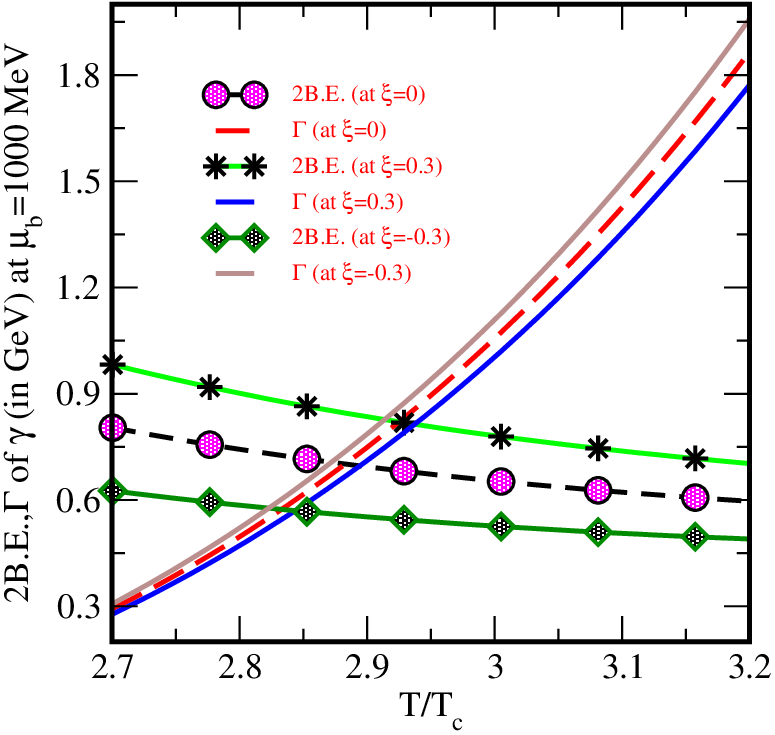}
    \vspace{2mm}
\caption{Shows the variation of 2B.E., $\Gamma$ of $\Upsilon$ with $T/T_{c}$ at different values of $\mu_{b}$ (left panel) and at different values of $\xi$ (right panel).}
\label{Fig.10}
 \vspace{3cm}  
\end{figure*}
%%%%%%%%%%%%%%%%%%%%%%%%%%%%%%%%%%%%%%%%%%%%%%%%%%%%%%%%%%%%%%%%%%%%%%%%%%%
%%%%%%%%%%%%%%%%%%%%%%%%%%%%%%%%%%%%%%%%%%%%%%%%%%%%%%%%%%%%%%%%%%%%%%%%%%%%%%%%%%%%%%
\begin{figure*}
    \vspace{2cm}   
    \includegraphics[height=6.5cm,width=6.5cm]{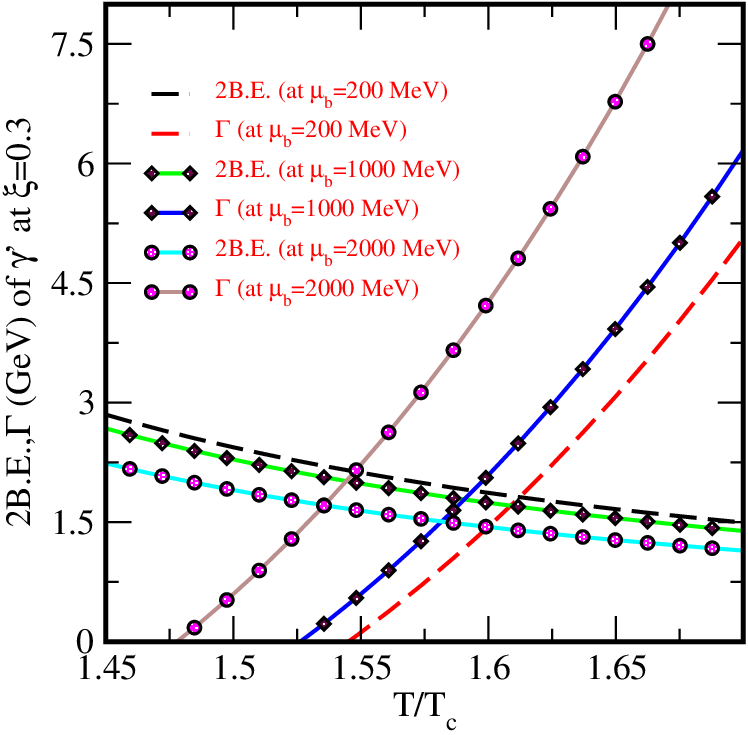}
    \hspace{2cm}
    \includegraphics[height=6.5cm,width=6.5cm]{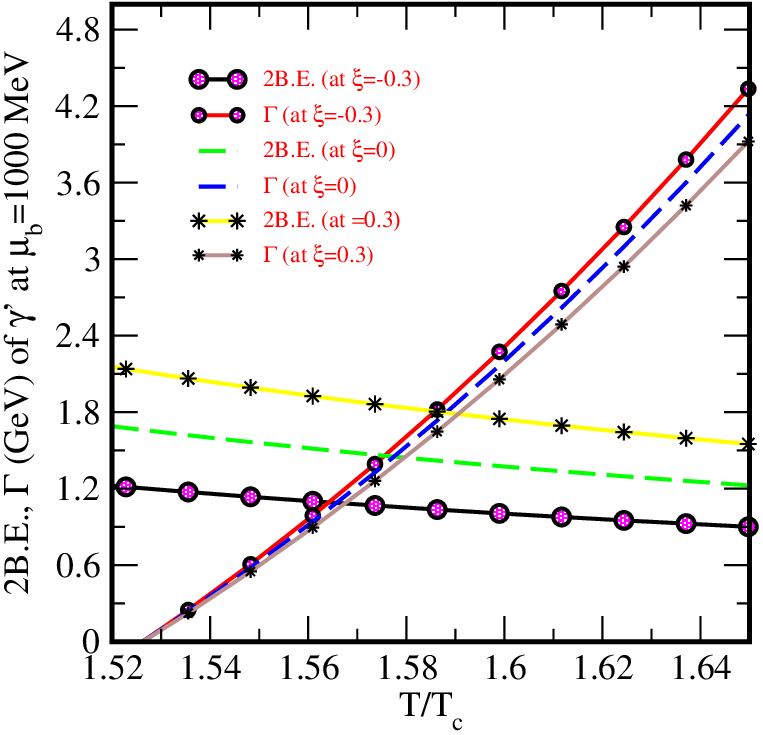}
\caption{Shows the variation of 2B.E., $\Gamma$ of $\Upsilon^{\prime}$ with $T/T_{c}$ at different values of $\mu_{b}$ (left panel) and at different values of $\xi$ (right panel).}
\label{Fig.11}
 \vspace{3cm}  
\end{figure*}
%%%%%%%%%%%%%%%%%%%%%%%%%%%%%%%%%%%%%%%%%%%%%%%%%%%%%%%%%%%%%%%%%%%%%%%%%%%%%%%%%%%%%%%%%%
%%%%%%%%%%%%%%%%%%%%%%%%%%%%%%%%%%%%%%%%%%%%%%%%%%%%%%%%%%%%%%%%%%%%%%%%%%%%%%%%%%%%%%
\begin{figure*}
    \vspace{2cm}   
    \includegraphics[height=6.5cm,width=6.5cm]{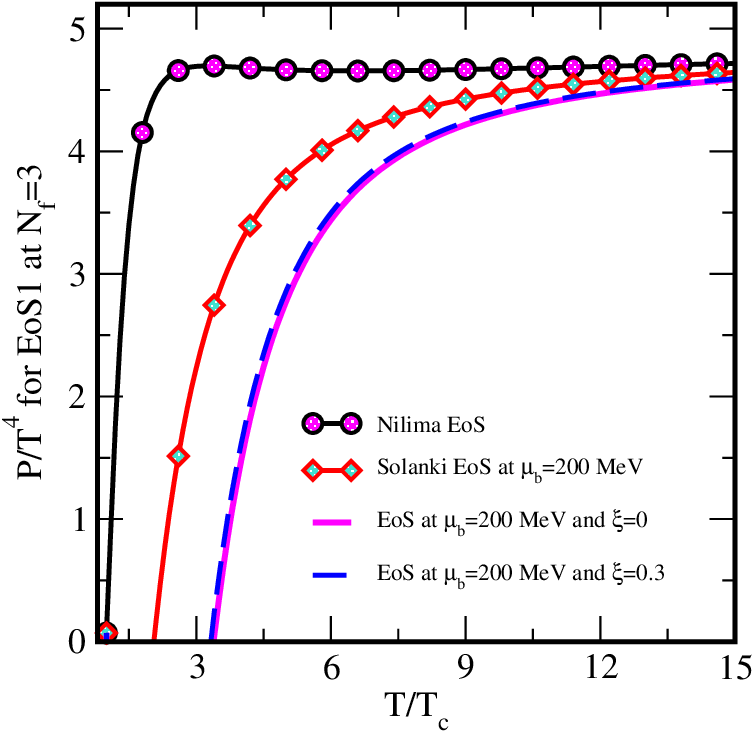}
    \hspace{2cm}
    \includegraphics[height=6.5cm,width=6.5cm]{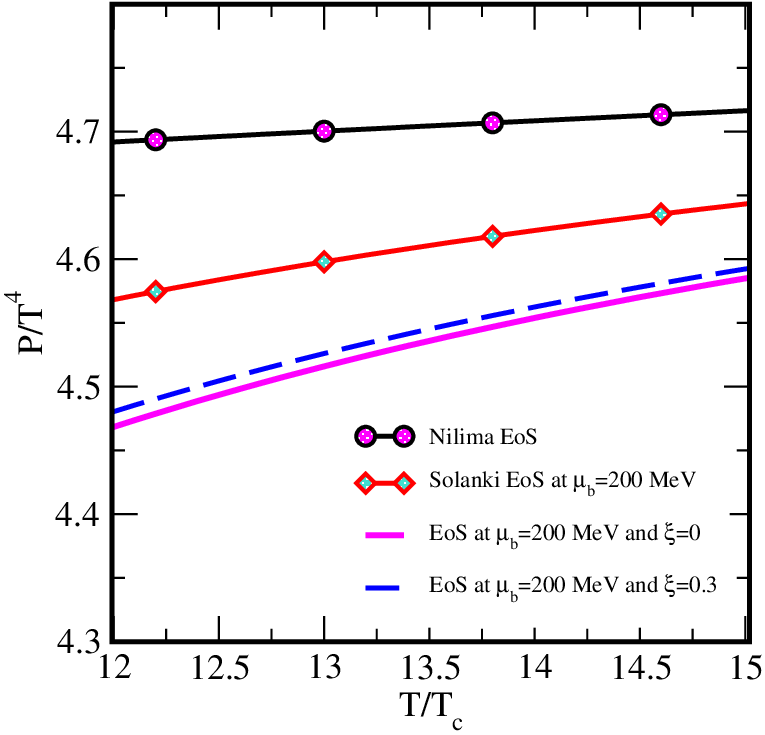}
\caption{Variation of P/$T^{4}$ with T/$T_{c}$ for EoS1 at $N_{f}$=3 quark-gluon plasma and potential is in parallel condition ($\theta$=0 degree) (left panel) and right panel figure shows the inner view of the minimum separation of left panel figure. In this figure black line with circle represents the results obtained from Nilima EoS's~\cite{87} and red line with diamond represents the results obtained from Solanki EoS's~\cite{39}.}
\label{Fig.12}
 \vspace{3cm}  
\end{figure*}
%%%%%%%%%%%%%%%%%%%%%%%%%%%%%%%%%%%%%%%%%%%%%%%%%%%%%%%%%%%%%%%%%%%%%%%%%%%%%%%%%%%%%%%%%%

\section{Binding energy (B.E.) of the different quarkonium s-states}
Now using the references~\cite{82,83,84}, binding energies of heavy quarkonium states in anisotropic medium can thus be obtained by solving Schrodinger equation with first order perturbation in anisotropy parameter ($\xi$). Now the expression of real part of binding energy is written below:
\begin{multline}
{\text{Re}[B.E]}=\frac{m_Q\sigma^2 }{m_{D}^{4}\left(T,\mu_{b}\right) n^{2}} + \alpha m_{D}\left(T,\mu_{b}\right)\\ +\frac{\xi}{3}\Big(\frac{m_Q\sigma^2 }{m_{D}^{4}\left(T,\mu_{b}\right) n^{2}}+ \alpha m_{D}\left(T,\mu_{b}\right)+ \frac{2m_Q\sigma^2 }{m_{D}^{4}\left(T,\mu_{b}\right) n^{2}}\Big).
\label{eq30}
\end{multline}
Where, n=1 and 2 corresponds to the ground and the first excited states of the heavy quarkonia respectively. It should be noted here that the above expression of binding energy(eq.~\ref{eq30}) is applicable only for $J/\psi$, $\Upsilon$, ${\Psi^{\prime}}$ and $ {\Upsilon^{\prime}}$. Fig.\ref{Fig.3} and \ref{Fig.4}, shows variation of the binding energy of $J/\psi$, $\Upsilon$, $\psi^{\prime}$ and $\Upsilon^{\prime}$ with T/$T_{c}$ at different values of baryonic chemical potential ($\mu_{b}$) (i.e., $\mu_{b}$=200, 1000 and 2000 MeV) at constant values of anisotropy ($\xi$) (i.e., $\xi$=0.3). From fig.\ref{Fig.3} and \ref{Fig.4} we have deduced that binding energy of $J/\psi$, $\Upsilon$, $\psi^{\prime}$ and $\Upsilon^{\prime}$ decreases if we increases the values of $\mu_{b}$.\\ 
Fig.\ref{Fig.5} and fig.\ref{Fig.6}, shows variation of the binding energy of $J/\psi$, $\Upsilon$, $\psi^{\prime}$ and $\Upsilon^{\prime}$ with T/$T_{c}$ at different values of anisotropy ($\xi$) (i.e., $\xi$=0.3, 0 and -0.3) at constant value of $\mu_{b}$ (i.e., $\mu_{b}$=1000 MeV). From fig.\ref{Fig.5} and \ref{Fig.6}, we have deduced that binding energy of $J/\psi$, $\Upsilon$, $\psi^{\prime}$ and $\Upsilon^{\prime}$ increases if we increases the values of $\xi$. We noticed that, the binding energy have higher values as one moves from prolate ($\xi<0$) to oblate ($\xi>0$) case. In anisotropic medium, the binding energy of $Q\bar{Q}$ pair get stronger with increase in anisotropy. This is due to the fact that, the variation of binding energy increases if we goes from prolate to oblate case, hence quarkonium states strongly bounds with anisotropy.\\ 

\section{Dissociation of quarkonium states in presence of $\xi$ and baryonic chemical potential}
The dissociation temperature for real binding energies can be obtained by using thermal energy effect. According to the references~\cite{85,86} it is not necessary to have zero binding energy for dissolution of the quarkonium states. When binding energy $(B.E. \le T$) of quarkonium state is weakly bonded, dissociates by means of thermal fluctuations. The quarkonium state also said to be dissociated when 
$2B.E.<\mathrm{\Gamma }(T)$, $\mathrm{\Gamma }(T)$ is thermal width of respective quarkonium states. When binding energy of charmonium and bottomonium state at a particular value of temperature becomes smaller or equal to the value of mean thermal energy, the state which said to be dissociated and this can eatimated by using the following expression B.E.=$T_{D}$ (for upper bound of quarkonium dissociation) and B.E.=$3T_{D}$ (for lower bound of quarkonium dissociation) as can be found in~\cite{Lal2023} and reference therein and is written as below:
\begin{multline}
B.E_{\left(J/\psi,\Upsilon, \psi^{'}, \Upsilon^{'}\mathrm{\ }\right)}= \frac{m_Q\sigma^2 }{m_{D}^{4}\left(T,\mu_{b}\right) n^{2}} + \alpha m_{D}\left(T,\mu_{b}\right)\\ +\frac{\xi}{3}\Big(\frac{m_Q\sigma^2 }{m_{D}^{4}\left(T,\mu_{b}\right) n^{2}}+ \alpha m_{D}\left(T,\mu_{b}\right)+ \frac{2m_Q\sigma^2 }{m_{D}^{4}\left(T,\mu_{b}\right) n^{2}}\Big)\\=\left\{ \begin{array}{cl}
T_{D} & : For \ Upper \ bound\\
3T_{D} & :For \ Lower \ bound
\end{array} \right.
\label{eq31}
\end{multline}
Further, we have calculated the dissociation temperature by using two criteria: firstly by using mean thermal energy and second by using thermal width. The dissociation temperature of quarkonium states by using mean thermal energy effect criteria are listed in the table \ref{tab.1} to table \ref{tab.4} for both lower and upper bound. In general, the dissociation temperature decreases with increase in the values of $\mu_{b}$ (i.e., $\mu_{b}$=200, 1000 and 2000 MeV), and increases with increase in the values of $\xi$ (i.e., $\xi$=-0.3, 0 and 0.3). 
%%%%%%%%%%%%%%%%%%%%%%%%%%%%%%%%%%%%%%%%%%%%%%%%%%%%%%%%%%%%%%%%%%%%%%%%%%%%%%%%%%%%%%%%%%%%%%%%%%%%%%%%%%%
\begin {table}
\caption {Lower bound of dissociation for $\xi$=0.3 at $T_{c}$=197 MeV}
 \label{tab.1} 
\begin{center}
\begin{tabular}{ |p{1cm}||p{2.3cm}|p{2.3cm}|p{2.3cm}|  }
\hline
  \multicolumn{4}{|c|}{Temperatures are in the unit of $T_c$}\\
 \hline
  \multicolumn{4}{|c|}{Dissociation by thermal energy effect criteria}\\
 \hline
 States $\Downarrow$ & $\mu_{b}$=200 MeV & $\mu_{b}$=1000 MeV & $\mu_{b}$=2000 MeV\\
 \hline
 $J/\psi$ & 1.6497 & 1.6243 & 1.5736\\
 $\Upsilon$ & 2.0558 & 2.0050 & 1.8908\\
 $\Upsilon^{\prime}$ & 1.6143 & 1.5931 & 1.5285\\
 \hline
 \end{tabular}
 \end{center}
\end{table}
%%%%%%%%%%%%%%%%%%%%%%%%%%%%%%%%%%%%%%%%%%%%%%%%%%%%%%%%%%%%%%%%%%%%%%%%%%%%%%%%%%%%%%%%%%%%%%%%%%%%%%%%%%%%
%%%%%%%%%%%%%%%%%%%%%%%%%%%%%%%%%%%%%%%%%%%%%%%%%%%%%%%%%%%%%%%%%%%%%%%%%%%%%%%%%%%%%%%%%%%%%%%%%%%%%%%%%%%
\begin {table}
\caption {Upper bound of dissociation for $\xi$=0.3 at $T_{c}$=197 MeV}
 \label{tab.2} 
\begin{center}
\begin{tabular}{ |p{1cm}||p{2.3cm}|p{2.3cm}|p{2.3cm}|  }
\hline
  \multicolumn{4}{|c|}{Temperatures are in the unit of $T_c$}\\
 \hline
  \multicolumn{4}{|c|}{Dissociation by thermal energy effect criteria}\\
 \hline
 States $\Downarrow$ & $\mu_{b}$=200 MeV & $\mu_{b}$=1000 MeV & $\mu_{b}$=2000 MeV\\
 \hline
 $J/\psi$ & 2.0812 & 2.0304 & 1.9162\\
 $\Upsilon$ & 2.6142 & 2.5253 & 2.3350\\
 $\Upsilon^{\prime}$ & 2.0431 &  1.9162 & 1.7893\\
 \hline
 \end{tabular}
 \end{center}
\end{table}
%%%%%%%%%%%%%%%%%%%%%%%%%%%%%%%%%%%%%%%%%%%%%%%%%%%%%%%%%%%%%%%%%%%%%%%%%%%%%%%%%%%%%%%%%%%%%%%%%%%%%%%%%%%%
%%%%%%%%%%%%%%%%%%%%%%%%%%%%%%%%%%%%%%%%%%%%%%%%%%%%%%%%%%%%%%%%%%%%%%%%%%%%%%%%%%%%%%
\begin{figure*}
    \vspace{2cm}   
    \includegraphics[height=6.5cm,width=6.5cm]{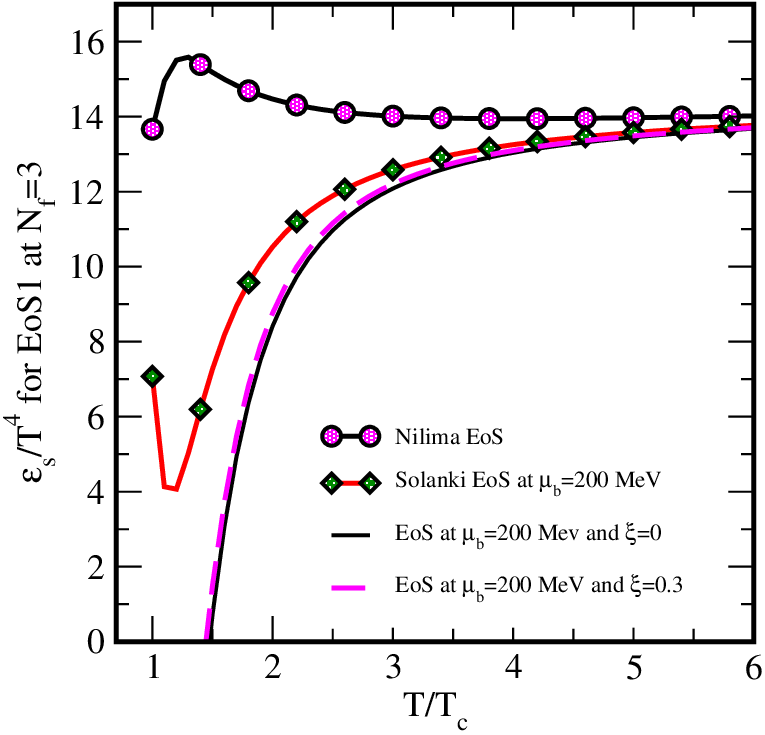}
    \hspace{2cm}
    \includegraphics[height=6.5cm,width=6.5cm]{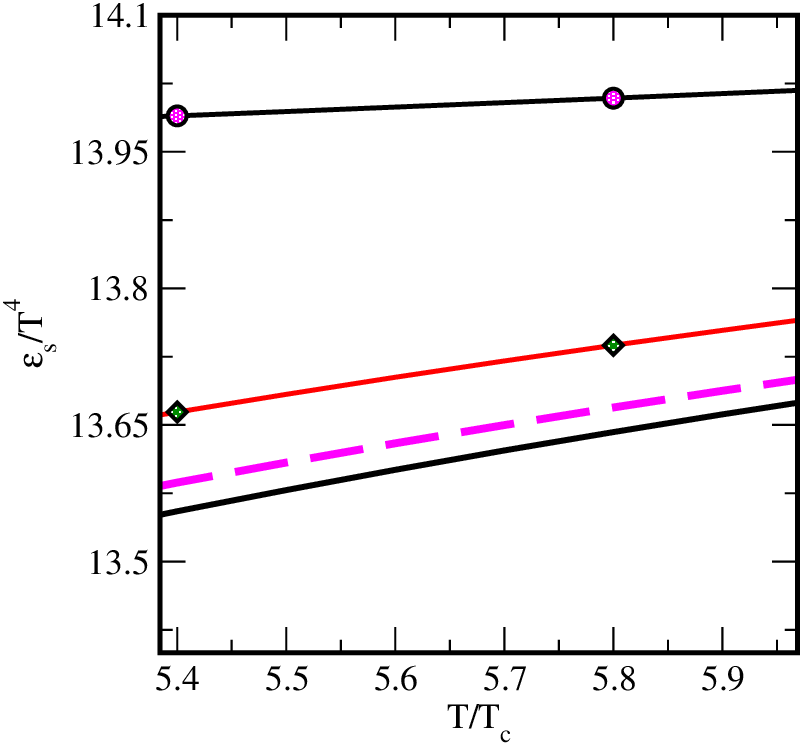}
\caption{Variation of $\epsilon_{s}/T^{4}$ with T/$T_{c}$ for EoS1 at $N_{f}$=3 quark-gluon plasma and potential is in parallel condition ($\theta$=0 degree) (left panel) and right panel figure shows the inner view of the minimum separation of left panel figure. In this figure black line with circle represents the results obtained from Nilima EoS's~\cite{87} and red line with diamond represents the results obtained from Solanki EoS's~\cite{39}.}
\label{Fig.13}
 \vspace{3cm}  
\end{figure*}
%%%%%%%%%%%%%%%%%%%%%%%%%%%%%%%%%%%%%%%%%%%%%%%%%%%%%%%%%%%%%%%%%%%%%%%%%%%%%%%%%%%%%%%%%%
%%%%%%%%%%%%%%%%%%%%%%%%%%%%%%%%%%%%%%%%%%%%%%%%%%%%%%%%%%%%%%%%%%%%%%%%%%%%%%%%%%%%%%%%%%%%%%%%%%%%%%%%%%%
\begin {table}
\caption {Lower bound of dissociation for $\mu_{b}$=1000 MeV at $T_{c}$=197 MeV}
 \label{tab.3} 
\begin{center}
\begin{tabular}{ |p{1cm}||p{2.3cm}|p{2.3cm}|p{2.3cm}|  }
\hline
  \multicolumn{4}{|c|}{Temperatures are in the unit of $T_c$}\\
 \hline
  \multicolumn{4}{|c|}{Dissociation by thermal energy effect criteria}\\
 \hline
 States $\Downarrow$ & $\xi$=-0.3 & $\xi$=0 & $\xi$=0.3\\
 \hline
 $J/\psi$ & 1.4593 & 1.5482 & 1.6243\\
 $\Upsilon$ & 1.7766 & 1.9035 & 2.0050\\
 $\Upsilon^{\prime}$ & 1.4086 & 1.5355 & 1.5931\\
 \hline
 \end{tabular}
 \end{center}
\end{table}
%%%%%%%%%%%%%%%%%%%%%%%%%%%%%%%%%%%%%%%%%%%%%%%%%%%%%%%%%%%%%%%%%%%%%%%%%%%%%%%%%%%%%%%%%%%%%%%%%%%%%%%%%%%%
\begin {table}
\caption {Upper bound of dissociation for $\mu_{b}$=1000 MeV at $T_{c}$=197 MeV}
 \label{tab.4} 
\begin{center}
\begin{tabular}{ |p{1cm}||p{2.3cm}|p{2.3cm}|p{2.3cm}|  }
\hline
  \multicolumn{4}{|c|}{Temperatures are in the unit of $T_c$}\\
 \hline
  \multicolumn{4}{|c|}{Dissociation by thermal energy effect criteria}\\
 \hline
 States $\Downarrow$ & $\xi$=-0.3 & $\xi$=0 & $\xi$=0.3\\
 \hline
 $J/\psi$ & 1.7766 & 1.9162 & 2.0304\\
 $\Upsilon$ & 2.2081 & 2.3730 & 2.5253\\
 $\Upsilon^{\prime}$ & 1.7893 &  1.8821 & 1.9162\\
 \hline
 \end{tabular}
 \end{center}
\end{table}
%%%%%%%%%%%%%%%%%%%%%%%%%%%%%%%%%%%%%%%%%%%%%%%%%%%%%%%%%%%%%%%%%%%%%%%%%%%%%%%%%%%%%%%%%%%%%%%%%%%%%%%%%%%%
%%%%%%%%%%%%%%%%%%%%%%%%%%%%%%%%%%%%%%%%%%%%%%%%%%%%%%%%%%%%%%%%%%%%%%%%%%%%%%%%%%%%%%%%%%%%%%%%%%%%%%%%%%%%
\begin {table}
\caption {Dissociation for $\mu_{b}$=1000 MeV at $T_{c}$=197 MeV}
 \label{tab.5} 
\begin{center}
\begin{tabular}{ |p{1cm}||p{2.3cm}|p{2.3cm}|p{2.3cm}|  }
\hline
  \multicolumn{4}{|c|}{Temperatures are in the unit of $T_c$}\\
 \hline
  \multicolumn{4}{|c|}{Dissociation by thermal width criteria}\\
 \hline
 States $\Downarrow$ & $\xi$=-0.3 & $\xi$=0 & $\xi$=0.3\\
 \hline
 $J/\psi$ & 1.3879 & 1.4202 & 1.4467\\
 $\Upsilon$ & 2.8232 & 2.8857 & 2.9409\\
 $\Upsilon^{\prime}$ & 1.5644 & 1.5788 & 1.5909\\
 \hline
 \end{tabular}
 \end{center}
\end{table}
%%%%%%%%%%%%%%%%%%%%%%%%%%%%%%%%%%%%%%%%%%%%%%%%%%%%%%%%%%%%%%%%%%%%%%%%%%%%%%%%%%%%%%%%%%%%%%%%%%%%%%%%%%%%
%%%%%%%%%%%%%%%%%%%%%%%%%%%%%%%%%%%%%%%%%%%%%%%%%%%%%%%%%%%%%%%%%%%%%%%%%%%%%%%%%%%%%%%%%%%%%%%%%%%%%%%%%%%
\begin {table}
\caption {Dissociation for $\xi$=0.3 at $T_{c}$=197 MeV}
 \label{tab.6} 
\begin{center}
\begin{tabular}{ |p{1cm}||p{2.3cm}|p{2.3cm}|p{2.3cm}|  }
\hline
  \multicolumn{4}{|c|}{Temperatures are in the unit of $T_c$}\\
 \hline
  \multicolumn{4}{|c|}{Dissociation by thermal width criteria}\\
 \hline
 States $\Downarrow$ & $\mu_{b}$=200 MeV & $\mu_{b}$=1000 MeV & $\mu_{b}$=2000 MeV\\
 \hline
 $J/\psi$ & 1.4618 & 1.4467 & 1.4082\\
 $\Upsilon$ & 3.0775 & 2.9385 & 2.6794\\
 $\Upsilon^{\prime}$ & 1.6127 & 1.5913 & 1.5379\\
 \hline
 \end{tabular}
 \end{center}
\end{table}
%%%%%%%%%%%%%%%%%%%%%%%%%%%%%%%%%%%%%%%%%%%%%%%%%%%%%%%%%%%%%%%%%%%%%%%%%%%%%%%%%%%%%%%%%%%%%%%%%%%%%%%%%%%%

\section{Thermal width of s-states of quarkonium}
As already mentioned in the section-III that, the quarkonia potential have both real and imaginary part. The real part gives rise to binding energy discussed earlier. Whereas, thermal width comes from imaginary part of the potential. The thermal width now employed for calculating dissociation point by exploiting twice of real binding energy with thermal width of the quarkonium states. Thus thermal width can be obtained as: 
%%%%%%%%%%%%%%%%%%%%%%%%%%%%%%%%%%%%%%%%%%%%%%%%%%%%%%%%%%%%%%%%%%%%%%%%%%%%%%%%%%%%%
\begin{figure*}
    \vspace{2cm}   
    \includegraphics[height=6.5cm,width=6.5cm]{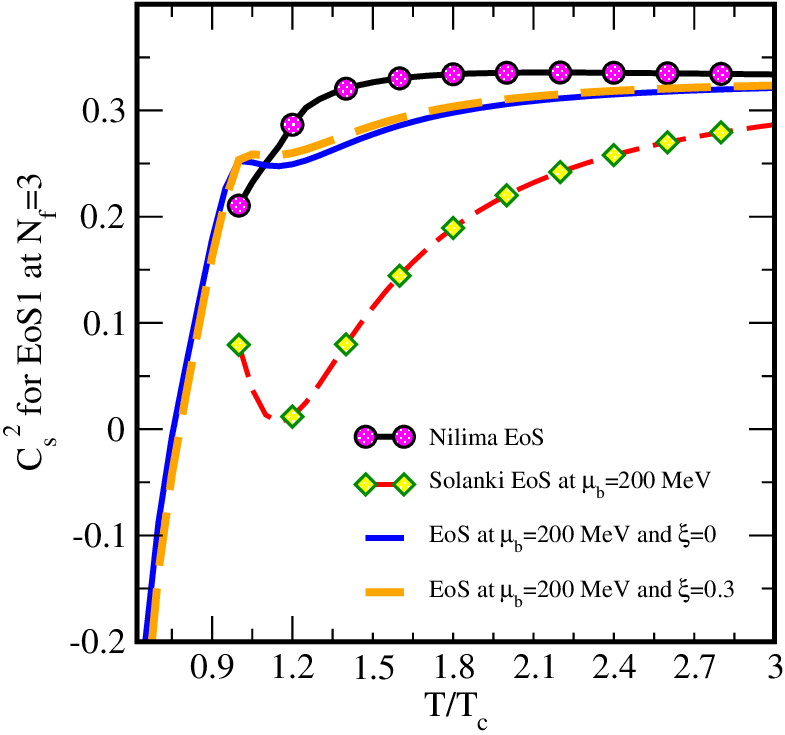}
    \hspace{2cm}
    \includegraphics[height=6.5cm,width=6.5cm]{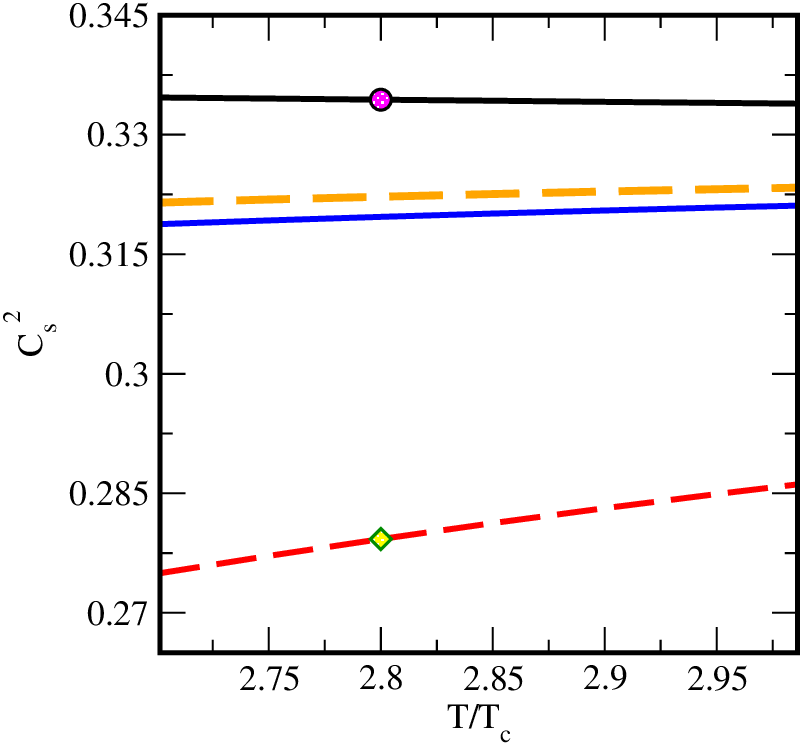}
\caption{Variation of $C_{s}^{2}$ with T/$T_{c}$ for EoS1 at $N_{f}$=3 quark-gluon plasma and potential is in parallel condition ($\theta$=0 degree) (left panel) and right panel figure shows the inner view of the minimum separation of left panel figure. In this figure black line with circle represents the results obtained from Nilima EoS's~\cite{87} and red line with bar represents the results obtained from Solanki EoS's~\cite{39}.}
\label{Fig.14}
 \vspace{3cm}  
\end{figure*}
%%%%%%%%%%%%%%%%%%%%%%%%%%%%%%%%%%%%%%%%%%%%%%%%%%%%%%%%%%%%%%%%%%%%%%%%%%%%%%%%%%%%%%%%%%
\ba
\Gamma(T) = - \int d^3{\bf{r}}\, \left|\Psi({{r}})\right|^2{\rm{Im}}~V({\bf{r}})
\label{eq32}
\ea
Where, $\Psi(r)$ is the Coulombic type wave function. The Coulombic wave function for J/$\psi$, $\Upsilon$, $\psi^{\prime}$ and $\Upsilon^{\prime}$ is given as:
\begin{eqnarray}
\Psi_{1S}(r)&=&\frac{1}{\sqrt{\pi a_0^3}}e^\frac{-r}{a_0}\nonumber\\
\Psi_{2S}(r)&=&\frac{1}{4\sqrt{2\pi a_{o}^{3}}}\left ( 2-\frac{r}{a_{o}} \right )e^{\frac{-r}{2a_{0}}}
\label{eq33}
\end{eqnarray}
Where, $a_0$=$2/(m_Q\alpha)$ represents Bohr radius of the quarkonia system. Now by using the eq.(\ref{eq33}), we have:
\begin{multline}
\Gamma_{1S/2S}(T)=m^{2}_{D}T\int r^{2}log\left(\frac{1}{r m_{D}}\right) d^{3}r|\Psi_{1s/2s}(r)|^{2}\\\left[ \frac{\alpha}{3} \left\{ \frac{\xi}{60}(7-9cos2\theta_{r})-1 \right\}\right]+m^{2}_{D}T\\\int r^{2}log\left(\frac{1}{r m_{D}}\right) d^{3}r|\Psi_{1s/2s}(r)|^{2}\\ \left[ \sigma r^{2}\left\{\frac{\xi}{35} \left(\frac{1}{9}-\frac{1}{4}cos2\theta_{r}\right)-\frac{1}{30} \right\} \right]
 \label{eq34}
\end{multline}
The thermal width for $1S$-state can be obtained by solving above equation as below:
\begin{multline}
\Gamma_{1S}(T)-\frac{m^{2}_{D}T(\xi-6)}{90\alpha^{4}m^{4}_{Q}}\\\left( 5(12\gamma-25)\alpha^{3}m^{2}_{Q}+9(20\gamma-49)\sigma \right)=\frac{m^{2}_{D}T(\xi-6)}{90\alpha^{4}m^{4}_{Q}}\\\left( 60(\alpha^{3}m^{2}_{Q}+3\sigma)log\left( \frac{\alpha M_{Q}}{m_{D}} \right) \right)
\label{eq35}
\end{multline}
Thus, the dissociation width for $1S$-state  up to leading logarithmic order of imaginary potential following reference~\cite{84}  would be of the form:
\ba
\frac{\Gamma_{1S}(T)}{m_D^2 \log\left(\frac{m_D}{\alpha m_Q}\right)}= T\bigg(\frac{4}{\alpha m_Q^2}+\frac{12\sigma}{\alpha ^4m_Q^4}\bigg) 
\bigg(1-\frac{\xi}{6}\bigg)
\label{eq36}
\ea 
Similarly for the 2S-state, using wave function for the 2S, state we have:
\begin{widetext}
\begin{multline}
\Gamma_{2S}(T)=\frac{T(\xi-6)}{45\alpha^{2}m_{Q}^{2}}m_{D}^{2}\left ( 35(12\gamma-31)\alpha)+\frac{72(160\gamma-447)\sigma}{\alpha^{2}m_{Q}^{2}}\right )+\frac{T(\xi-6)}{45\alpha^{2}m_{Q}^{2}}m_{D}^{2}\left \{60\left ( 7\alpha+\frac{192\sigma}{{\alpha^{2}m_{Q}^{2}}} \right )log\frac{\alpha m_{Q}}{2m_{D}}\right\}
\label{eq37}
\end{multline}
\end{widetext}
And the leading logarithmic order for 2S-state is given as:
\begin{multline}
\frac{\Gamma_{2S}(T)}{log\left ( \frac{2m_{D}}{\alpha m_{Q}}\right )}=\frac{8m_{D}^{2}T}{\alpha^{4}m_{Q}^{4}}\left ( 1-\frac{\xi}{6} \right )(7\alpha^{3}m_{Q}^{2} + 192\sigma)
\label{eq38}
\end{multline}
The dissociation temperature of different quarkonium states by exploiting the thermal width and twice of real binding energy has been shown in the fig.\ref{Fig.9} (for J/$\psi$), fig.\ref{Fig.10} (for $\Upsilon$) and fig.\ref{Fig.11} (for $\Upsilon^{\prime}$). The dissociation temperature obtained from the intersection point of twice of real binding energy and thermal width for different states at different values of anisotropy ($\xi$) and baryonic chemical potential are enlisted in the tables \ref{tab.5} and \ref{tab.6}. There were no dissociation temperature found for the $\psi^{\prime}$ due to its small mass value and hence decays earlier than ground state.
%%%%%%%%%%%%%%%%%%%%%%%%%%%%%%%%%%%%%%%%%%%%%%%%%%%%%%%%%%%%%%%%%%%%%%%%%%%%%%%%%%%%%%%%%%%%%%%%%%%%%%%%%%%
\begin{table*}
\centering
\caption {Mass spectra of ground state of quarkonium at $\xi$=0}
 \label{tab.7} 
\begin{tabular}{ |p{1cm}||p{2cm}|p{2cm}|p{2cm}|p{2cm}|p{2cm}|  }
\hline
  \multicolumn{6}{|c|}{Mass spectra are in the unit of GeV}\\
 \hline
  \multicolumn{4}{|c|}{For $m_{J/\psi}$=1.5 GeV and $m_{\Upsilon}$=4.5 GeV}\\
 \hline
 States $\Downarrow$ & $\mu_{b}$=200 MeV & $\mu_{b}$=1000 MeV & $\mu_{b}$=2000 MeV & Theoritical Result~\cite{39} & Experimental Result~\cite{96}\\
 \hline
 $J/\psi$ & 3.520 & 3.480 & 3.391 & 3.060 & 3.096\\
 $\Upsilon$ & 10.32 & 10.18 & 9.909 & 9.200 & 9.460\\
 \hline
 \end{tabular}
\end{table*}
%%%%%%%%%%%%%%%%%%%%%%%%%%%%%%%%%%%%%%%%%%%%%%%%%%%%%%%%%%%%%%%%%%%%%%%%%%%%%%%%%%%%%%%%%%%%%%%%%%%%%%%%%%%%

\section{Mass Spectra of quarkonium states in the presence of anisotropy and baryonic chemical potential}
The mass spectra of $1S$ and $2S$ states of charmonium and bottomonium in anisotropic medium can be calculated by using following conditions:
\ba
M=2m_{Q}+B.E 
\label{eq39}
\ea  
Hence, we have:
\begin{multline}
\text{Mass spectra of quarkonium states}=\\ 2m_{Q} + \bigg( \frac{m_Q\sigma^2 }{m_{D}^4 n^{2}} + \alpha m_{D} +\frac{\xi}{3}\Big(\frac{m_Q\sigma^2 }{m_{D}^4 n^{2}}+ \alpha m_{D} + \frac{2m_Q\sigma^2 }{m_{D}^4 n^{2}}\Big)\bigg)
\label{eq40}
\end{multline}             
Where, $m_Q$ is the mass of heavy quarkonia.\\
Fig.\ref{Fig.7} and fig.\ref{Fig.8}, shows the variation of mass spectra of $J/\psi$ (left panel) and $\Upsilon$ (right panel) with T/$T_{c}$ at different values of baryonic chemical potential ($\mu_{b}$) (i.e., $\mu_{b}$=200, 1000 and 2000 MeV) (in fig.\ref{Fig.7}) and at different values of $\xi$ (i.e., $\xi$=-0.3, 0 and 0.3) (in fig.\ref{Fig.8}). From Fig.\ref{Fig.7} and \ref{Fig.8} it was deduced that mass spectra of $J/\psi$ and $\Upsilon$ decreases if we increases the values of $\mu_{b}$ and increases if we increases the values of $\xi$. In table VII and table VIII, we have calculated the values of mass spectra. In table-VII, we noticed that if we increases the values of $\mu_{b}$, then the values of mass spectra decreases. In table VIII, we noticed that if we increases the values of $\xi$ then the values of mass spectra also increases. We have also compared the results of mass spectra at different values of $\mu_{b}$ (in table VII) and at different values of $\xi$ (in table VIII) with the previous published theoretical~\cite{39} and experimental~\cite{96} results. From table we have seen that, the values of this result of mass spectra is approximately near to the experimental and theoretical result value. 

\section{Thermodynamical properties of quark-matter with anisotropic parameter ($\xi$) using EoS's of QGP}
The EoS's played invaluable role to understand the behavior of the QGP which is produced in the relativistic nucleus-nucleus collisions. EoS's are very sensitive to the matter and are important to verify/investigates the quarkonium suppression~\cite{89,90}. The expansion of QGP is highly sensitive to EoS's via the speed of sound, and it investigates the sensitivity of quarkonium suppression to the EoS~\cite{89,90}. Bannur~\cite{88} created an EoS for a strongly coupled QGP by appropriate modification of strongly coupled plasma in QED with integrating the running coupling constant and making suitable adjustment to account for color and flavor degree of freedom, and found a pretty good fit to the lattice findings. Now, we have gone through the EoS's, which is stated as a function of plasma parameter~\cite{91} briefly:
\begin{eqnarray}
\label{eq41}
\epsilon_{QED}-nT\mu_{ex}(\Gamma )=\frac{3}{2}nT
\end{eqnarray}
Where, the first term represent the ideal contribution and the deviations from ideal EoS as given below:
\begin{multline}
\label{eq42}
\mu_{ex}(\Gamma)(1+3\times10^{3}\Gamma^{5.7})\\=\mu^{Abe}_{ex}(\Gamma)+3\times10^{3}\Gamma^{5.7}\mu^{OCP}_{ex}(\Gamma)
\end{multline}
Where, $\mu^{Abe}_{ex}$ is: 
\begin{eqnarray}
\label{eq43}
\mu^{Abe}_{ex}+3\Gamma^{3}\left[ \frac{3}{8}ln(3\Gamma)+\frac{\gamma}{2}-\frac{1}{3} \right]=-\frac{\sqrt{3}}{2}\Gamma^{\frac{3}{2}}
\end{eqnarray}
Where the term $\mu^{Abe}_{ex}$, determined for plasma component and valid for all $\Gamma<180$ ~\cite{92}, is given below:
Where, the term $\mu^{OCP}_{ex}$ is,
\begin{multline}
\label{eq44}
\mu^{OCP}_{ex}-(0.220703\Gamma^{-\frac{1}{4}}-0.86097)\\=-0.898004\Gamma+0.96786\Gamma^{\frac{1}{4}}
\end{multline}
For strongly coupled plasma, in QCD it was assumed that the hadron exists for $T<T_{c}$ and goes to QGP for $T>T_{c}$. At $T>T_{c}$,
it is the strongly interacting plasma of quarks, gluons and no hadrons because it was assumed that interaction of confinement due to QCD vacuum has been melted~\cite{88} at $T=T_{c}$. Hence, only coulomb interaction present in the deconfined plasma phase. So, the plasma parameter, which is the ratio of particle average potential energy to particle average kinetic energy, is assumed to be weak $\Gamma \ll 1$ and as given by:
\begin{eqnarray}
\label{eq45}
\Gamma\equiv \frac{<PE>}{<KE>}= \frac{Re[V(r,T)]}{T}
\end{eqnarray}
Finally, the EoS has been obtained by using the potential eq.(\ref{eq8}) in the plasma parameter after inclusion of quantum and relativistic effects as:
\begin{eqnarray}
\label{eq46}
\frac{\epsilon_{s}}{nT} =\left ( 3+\mu_{ex}(\Gamma) \right )
\end{eqnarray}
Where, the $\mu_{ex}$ remains same as in eq.(\ref{eq42}). The scaled energy density is now expressed in terms of ideal contribution:
\begin{eqnarray}
\label{eq47}
e(\Gamma)\equiv \frac{\epsilon_{s} }{\epsilon_{SB } }=1+\frac{1}{3}\mu_{ex}(\Gamma)
\end{eqnarray}
\begin{eqnarray}
\label{eq48}
\epsilon_{SB}\equiv (16+21N_{f}/2)\pi^{2}T^{4}/30
\end{eqnarray}\\
Here, $N_{f}$ denotes the number of quark and gluon flavors. For the $\overline{MS}$ approach, we have now used two loop level QCD running coupling constants~\cite{93}.
\begin{eqnarray}
\label{eq49}
g^{2}(T)\approx 2b_{0}ln\frac{\bar{\mu}}{\Lambda_{\overline{MS}} }\left ( 1+\frac{b_{1}}{2b_{0}^{2}}\frac{ln\left ( 2ln\frac{\bar{\mu}}{\Lambda_{\overline{MS}} } \right )}{ln\frac{\bar{\mu}}{\Lambda_{\overline{MS}} }} \right )^{-1}
\end{eqnarray}
Where, $b_{0}=\frac{33-2N_{f}}{48\pi^{2}}$ and $b_{1}=\frac{153-19N_{f}}{384\pi^{4}}$. In case of $\overline{MS}$ scheme, $\Lambda_{\overline{MS}}$ and $\bar{\mu}$ are considered as the renormalization scale and the scale parameter respectively:
\begin{eqnarray}
\label{eq50}
\bar{\mu}exp(\gamma_{E}+c)=\Lambda _{\overline{MS}}(T)\nonumber\\
\Lambda _{\overline{MS}}(T)exp(\gamma_{E}+c)=4\pi\Lambda_{T}.
\end{eqnarray}
%%%%%%%%%%%%%%%%%%%%%%%%%%%%%%%%%%%%%%%%%%%%%%%%%%%%%%%%%%%%%%%%%%%%%%%%%%%%%%%%%%%%%%%%%%%%%%%%%%%%%%%%%%%
\begin{table*}
\centering
\caption {Mass spectra of ground state of quarkonium at $\mu_{b}$=1000 MeV}
 \label{tab.8} 
\begin{tabular}{ |p{1cm}||p{2cm}|p{2cm}|p{2cm}|p{2cm}|p{2cm}|  }
\hline
  \multicolumn{6}{|c|}{Mass spectra are in the unit of GeV}\\
 \hline
  \multicolumn{4}{|c|}{For $m_{J/\psi}$=1.5 GeV and $m_{\Upsilon}$=4.5 GeV}\\
 \hline
 States $\Downarrow$ & $\xi$=-0.3 & $\xi$=0 & $\xi$=0.3 & Theoritical Result~\cite{39} & Experimental Result~\cite{96}\\
 \hline
 $J/\psi$ & 3.361 & 3.480 & 3.597 & 3.060 & 3.096\\
 $\Upsilon$ & 9.864 & 10.18 & 10.53 & 9.200 & 9.460\\
 \hline
 \end{tabular}
\end{table*}
%%%%%%%%%%%%%%%%%%%%%%%%%%%%%%%%%%%%%%%%%%%%%%%%%%%%%%%%%%%%%%%%%%%%%%%%%%%%%%%%%%%%%%%%%%%%%%%%%%%%%%%%%%%%
\begin{figure*}
    \vspace{5cm}   
    \includegraphics[height=4.5cm,width=5cm]{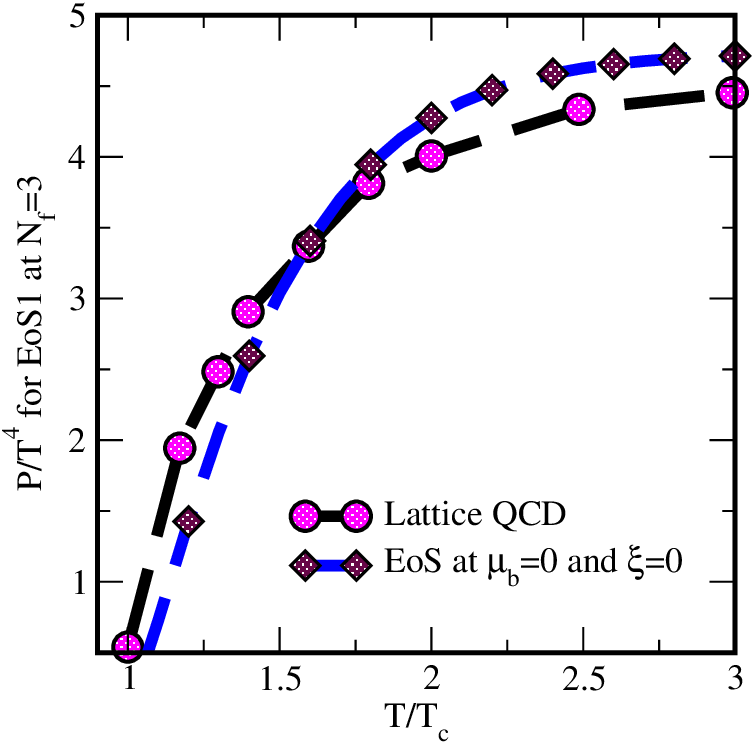}
    \hspace{1cm}
    \includegraphics[height=4.5cm,width=5cm]{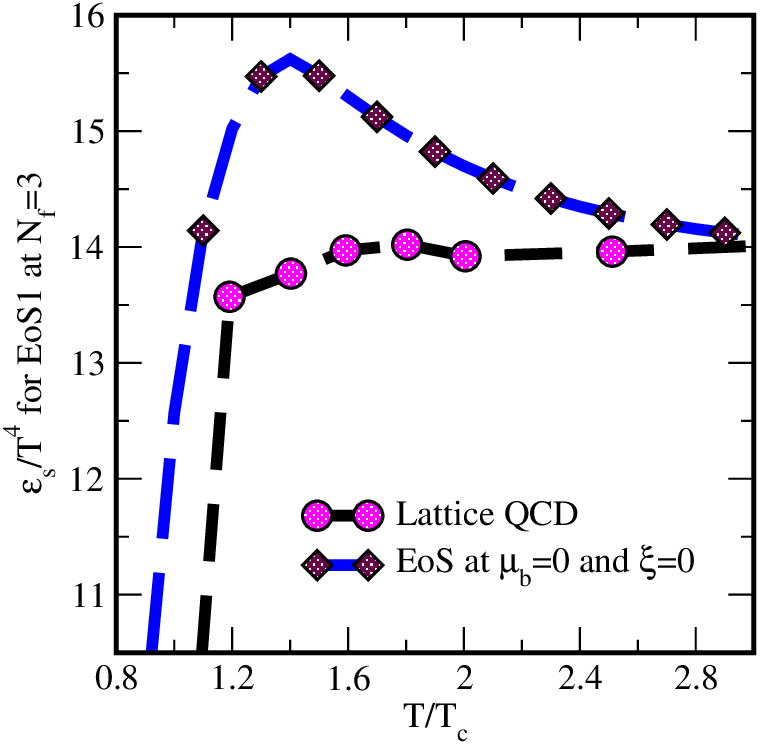}
    \hspace{1cm}
    \includegraphics[height=4.5cm,width=5cm]{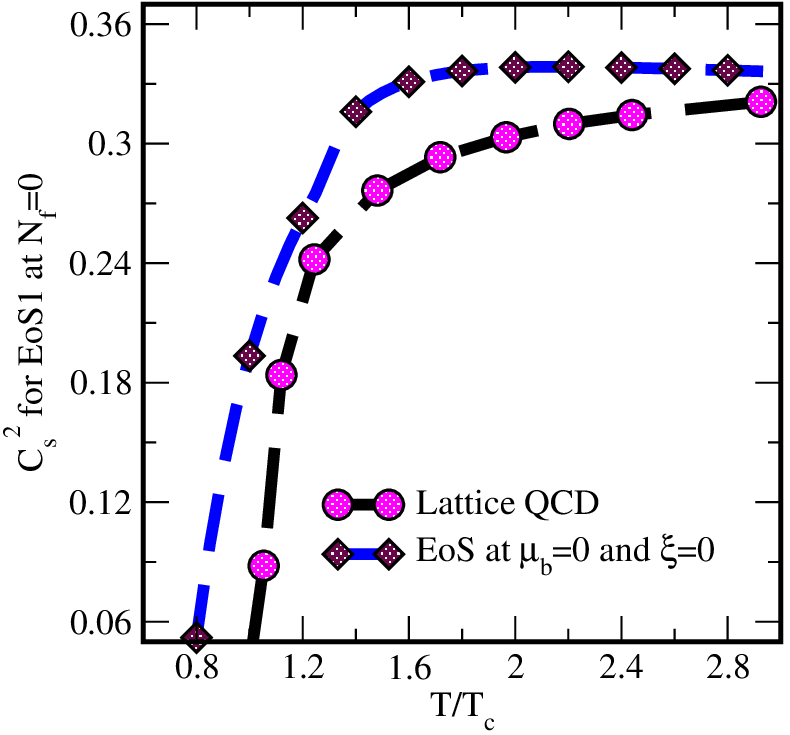}
\caption{Variation of P/$T^{4}$ with T/$T_{c}$ (left panel), $\epsilon_{s}/T^{4}$ with T/$T_{c}$ (middle panel)for EoS1 at $N_{f}$=3 and $C_{s}^{2}$ with T/$T_{c}$ (right panel) for EoS1 at $N_{f}$=0 quark-gluon plasma and potential is in parallel condition ($\theta$=0 degree). In this figure black line with circles represents the lattice QCD results (for pure gauge) obtained from~\cite{87} and blue line with diamonds represents the Our EoS at $\xi$=0 and $\mu_{b}$=0.}
\label{Fig.15}
 \vspace{2mm}  
\end{figure*}
%%%%%%%%%%%%%%%%%%%%%%%%%%%%%%%%%%%%%%%%%%%%%%%%%%%%%%%%%%%%%%%%%%%%%%%%%%%%%%%%%%%%%%%%%%
Here, $\gamma_{E}$=0.5772156 and $c=\frac{N_{c}-4N_{f}ln4}{22N_{c}-N_{f}}$ that is a constant depending on colors and ﬂavors. There are various uncertainties in the formula for the running coupling constant, which are connected with the scale parameter and renormalization scale $\overline{MS}$. This problem has been superseded by using Brodsky, Lepage and Mackenzie criteria ~\cite{94}. $\overline{MS}$ was permitted to fluctuate between $\pi$T and 4$\pi$T~\cite{95}. For our motivation, we chose the $\overline{MS}$ near to the center value 2$\pi$$T_{c}$~\cite{86} for $N_{f}$=0, and $T_{c}$ for both $N_{f}$=2 and $N_{f}$=3 flavors. When the factor $\frac{b_{1}}{2b_{0}^{2}}\frac{ln\left ( 2ln\frac{\bar{\mu}}{\Lambda_{\overline{MS}} } \right )}{ln\frac{\bar{\mu}}{\Lambda_{\overline{MS}} }}$ is $\gg$ 1 then expression is reduced as used in~\cite{88}, ignoring the higher order terms of the preceding component. However, this option does not hold true for the temperature ranges employed in the computation, resulting in a coupling mistake that ultimately causes the difference in findings between our model and the Bannur model~\cite{88}. First, we have computed the energy density $\epsilon_{s}$(T) using eq.(\ref{eq47}) and the thermodynamic relation:
\begin{eqnarray}
\label{eq51}
\epsilon_{s}+P =T\frac{dp}{dT}
\end{eqnarray}
Further, pressure was calculated as:
\begin{eqnarray}
\label{eq52}
\frac{P}{T^{4}}=\left ( \frac{P_{0}}{T_{0}}+3a_{f}\int_{T_{0}}^{T}d\tau \tau^{2}e(\Gamma (\tau ))  \right )/T^{3}
\end{eqnarray}

Here, $P_{0}$ denotes the pressure at some temperature $T_{0}$ and ${a_f}=(16+\frac{21}{2}$${N_f})$$\frac{\pi^2}{90}$$T^4 $. Thus, speed of sound that can be evaluated once we have  pressure (P) and energy density ($\epsilon_{s}$) in hand and is given below:
\begin{eqnarray}
\label{eq53}
c_{s}^{2}d\epsilon_{s}=dP
\end{eqnarray}
All the above thermodynamical properties is potential dependent, and the potential is Debye mass dependent. In that case, we invade the problem by trading off the dependence on baryonic chemical potential ($\mu_{b}$), anisotropy ($\xi$) and temperature to a dependence on these thermodynamic properties of matter.\\
The thermodynamical properties of quark matter (i.e., pressure, energy density and speed of sound) plays a curious role in the study of QGP and also provide useful information about the strange quark-matter. The thermodynamic behavior of QCD matter at high temperature or above critical temperature is currently studied by lattice QCD~\cite{S.Plumari,H.Berrehrah}. In fig.(\ref{Fig.12}) we have plotted the variation of pressure $(\frac{P}{T^{4}})$ with temperature $(T/T_{c})$ using EoS1 for $N_{f}$=3 QGP along with Nilima EoS's~\cite{87} and Solanki EoS's~\cite{39}. Now, energy density $\epsilon_{s}$, speed of sound $(C_{s}^{2})$, and so forth can be derived since we had obtained pressure. In fig.(\ref{Fig.13}), we have plotted the energy density ($\frac{\epsilon_{s}}{T^{4}}$) with temperature $(T/T_{c})$ using Eos1 for $N_{f}$=3 QGP along with Nilima EoS's~\cite{87} and Solanki EoS's~\cite{39}. In figure (\ref{Fig.14}), we have plotted the speed of sound $(C_{s}^{2})$ with temperature ($T/T_{c}$) using Eos1 for $N_{f}$=3 QGP along with Nilima EoS's~\cite{87} and Solanki EoS's~\cite{39}. Our results of these thermodynamical properties of quark matter is approximately matched with the result of Nilima EoS's~\cite{87} and Solanki EoS's~\cite{39} with anisotropy parameter. The effect of anisotropy was also observed in these thermodynamical properties of quark matter as shown in figures (\ref{Fig.12}), (\ref{Fig.13}) and (\ref{Fig.14}). If we increases the value of anisotropy ($\xi$=0 to 0.3), then the variation of P/$T^{4}$, $\epsilon_{s}$/$T^{4}$ and $C_{s}^{2}$ also increases slightly respectively (the right panel of the figures \ref{Fig.12}, \ref{Fig.13} and \ref{Fig.14} is shows the minimum separation of the left panel). In figure (\ref{Fig.15}), we have shows the variation of $(\frac{P}{T^{4}})$ with temperature ($T/T_{c}$) (left panel), $\epsilon_{s}$/$T^{4}$ with temperature ($T/T_{c}$) (middle panel) using EoS1 at $N_{f}$=3 and $(C_{s}^{2})$ with temperature ($T/T_{c}$) (right panel) using EoS1 at $N_{f}$=0, and compared with the lattice QCD results~\cite{87,88}. Since lattice QCD (LQCD) results are available for only pure gauge, therefore comparison (in figure (\ref{Fig.15})) has been taken for the above mentioned value of flavor $N_{f}$ only. Our flavored results match approximately good with the LQCD results at $\xi$=0 and $\mu_{b}$=0. The main featured are the sharp rise of the curves of $(\frac{P}{T^{4}})$, $\epsilon_{s}$/$T^{4}$ and $(C_{s}^{2})$ around the value of critical temperature and then shows a linear curve to the ideal value. We calculate, these thermodynamical properties (i.e., $(\frac{P}{T^{4}})$, $\epsilon_{s}$/$T^{4}$ and $(C_{s}^{2})$) to calculate hydrodynamical expansion of Quark Gluon Plasma, and in future we extend our work to calculate the suppression of quarkonia in nuclear collisions with the effect of anisotropy and baryonic chemical potential.

\section{Results and Conclusions}
This work is devoted to study the effect of baryonic chemical potential ($\mu_{b}$) on quarkonia properties in anisotropic medium, by considering complex potential having both perturbative and non-perturbative nature, using quasi particle Debye mass. It is known that, anisotropy arises in primary stages as system expands after Ultra-Relativistic Heavy Ion Collisions (URHIC's) process. At condition $\xi$=0, string term ($\sigma$) of Cornell potential makes potential more attractive. This leads to the fact that, respective quarkonium states becomes more bound in comparison to the case when Coulombic term of potential modified alone. In this work, we have considered the values of anisotropy for three case viz. prolate ($\xi$=-0.3), isotropic ($\xi$=0) and oblate ($\xi$=0.3) with fixed value of critical temperature $T_{c}$=197 MeV.\\
We have reconsidered the medium modified form of heavy quark potential at finite values of $\mu_{b}$ and $\xi$. This has been done by considering real and imaginary part of potential with static gluon propagator which in turn gives the real and imaginary part of dielectric permittivity with anisotropic parameter. We considered $\mu_{b}$ and temperature dependent quasi-particle Debye mass to the study of dissociation pattern of quarkonia. Real part of potential has been used for solving schrodinger equation to obtain binding energy of quarkonia, and the imaginary part give rise to thermal width of heavy quarkonia. We observed that binding energy decreases and thermal width increases with increasing the values of $\mu_{b}$. However, binding energy tends to get higher with increasing value of $\xi$. In conclusion, the dissociation temperature of heavy quarkonia decreases with baryonic chemical potential and increases with the anisotropy as shown in table I to table VI.\\ 
We have also calculated the values of mass spectra, and noticed that if we increases the values of $\mu_{b}$ then the values of mass spectra decreases, but if we increases the values of $\xi$ then the values of mass spectra also increases. We also extend this work to calculate the thermodynamical properties of the QGP with $\xi$ and $\mu_{b}$.These EoS's are important to study the Suppression phenomena in the presence of $\xi$ and $\mu_{b}$. We have also extended this work, after calculating the thermodynamical properties of QGP (i.e., pressure, energy density and speed of sound) using the $\xi$ and $\mu_{b}$, mainly for the calculation of nucleus-nucleus suppression with the effect of anisotropy and baryonic chemical potential. We found that, if we increases the values of $\xi$ from 0 to 0.3, variation of pressure, energy density and speed of sound with T/$T_{c}$ increases little bit.\\
In future, we will extend this work to calculate the Survival property of different quarkonium states in the presence of $\xi$ and $\mu_{b}$ at different states of energy density ($\sqrt{s_{NN}}$), this survival probability will be calculated with respect to anisotropy, baryonic chemical potential, transverse momentum, centrality, and rapidity which is the key point to quantify various properties of the medium produced during  Heavy Ion Collisions (HICs) at LHC and RHIC. The results of this work might be helpful for expanding the studies of the highly dense objects like neutron stars. Since the Compressed Baryonic Matter (CBM) experiment at facility for anti proton and ion research (FAIR) is exploring the QGP at higher baryon densities, so such type of theoretical studies may participate to the physics of highly dense bodies with high baryon densities. 

\subsection{Acknowledgement}
VKA acknowledge the Science and Engineering research Board (SERB) Project No. {\bf EEQ/2018/000181} New Delhi for providing the financial support. We record our sincere gratitude to the people of India for their generous support for the research in basic sciences.

\subsection{Appendix}
Advantageous representation of propagators in the real-time formalism is the Keldysh representation where the 4-components of the matrix form is in linear combination, among these 4-components of the matrix, 3-components are independent, give the relation for advanced (A), retarded (R) and symmetric (F) propagators respectively:
\begin{align}
D_{R}^{0}=D_{11}^{0}-D_{12}^{0},\: D_{A}^{0}=D_{11}^{0}-D_{21}^{0},\: D_{F}^{0}=D_{11}^{0}-D_{22}^{0}\tag{A}
\end{align}
In the distribution function only F-components involves and particularly useful for the HTL diagrams. Similar relations for the self energies are:
\begin{align}
\Pi_{R}=\Pi_{11}+\Pi_{12},\: \Pi_{A}=\Pi_{11}+\Pi_{21},\: \Pi_{F}=\Pi_{11}+\Pi_{22}\tag{B}
\end{align}
Resumming the Dyson-Schwinger equation, the R, A and F propagators can be written as:
\begin{align}
D_{R,A}=D_{R,A}^{0}+D_{R,A}^{0}\Pi_{R,A}D_{R,A}\tag{C}
\end{align}
and
\begin{align}
D_{F}=D_{F}^{0}+D_{R}^{0}\Pi_{R}D_{F}+D_{F}^{0}\Pi_{A}D_{A}+D_{R}^{0}\Pi_{F}D_{A}\tag{D}
\end{align}
Now place the F-propagator $D_{F}^{0}(P)$ in terms of R and A propagators, the resummed F propagators is:
\begin{widetext}
\begin{align}
D_{F}(P)=(1+2f_{B})sgn(p_{0})[D_{R}(P)-D_{A}(P)]+D_{R}(P)[\Pi _{F}(P)-(1+2f_{B})sgn(p_{0})[\Pi _{R}(P)-\Pi _{A}(P)]]D_{A}(P)\tag{E}
\end{align}
\end{widetext}
For the calculation of static potential in $\xi$=0 medium, only the temporal (L) component of the propagator is requaired, so the R and A propagator in the form of simplest coulomb gauge is:
\begin{align}
D_{R,A(iso)}^{L}=D_{R,A}^{L(0)}+D_{R,A}^{L(0)}\Pi _{R,A(iso)}^{L}D_{R,A(iso)}^{L}\tag{F}
\end{align}
Now, we first enhance the self energy and propagators around $\xi$=0 limit and withhold only the linear term:
\begin{align}
D=D_{iso}+\xi D_{aniso},\; \Pi =\Pi _{iso}+\xi \Pi _{aniso}\tag{G}
\end{align}
The L component of the R and A propagatorin the presence of small $\xi$ becomes,
\begin{widetext}
\begin{align}
D_{R,A(aniso)}^{L}=D_{R,A}^{L(0)}\Pi_{R,A(aniso)}^{L}D_{R,A(iso)}^{L}+D_{R,A}^{L(0)}\Pi_{R,A(iso)}^{L}D_{R,A(aniso)}^{L}\tag{H}
\end{align}
\end{widetext}
Where as the notations for the difference of self energies and the propagators can be obtained~\cite{47,12}. For the solution of propagators, now we will calculate the gluon self energy for the gluon and quark loops~\cite{12} with external and internal momenta respectively with $Q=K-P$:
\begin{align}
\Pi^{\mu\nu}(P)=-\frac{i}{2}N_{f}g^{2}\int \frac{d^{4}K}{(2\pi)^{4}}tr[\gamma^{\mu}S(Q)\gamma^{\nu}S(K)]\tag{I}
\end{align}
And the R self energy is:
\begin{widetext}
\begin{align}
\Pi_{R}^{\mu\nu}(P)=-\frac{i}{2}N_{f}g^{2}\int \frac{d^{4}K}{(2\pi)^{4}}(tr[\gamma^{\mu}S_{11}(Q)\gamma^{\nu}S_{11}(K)]-tr[\gamma^{\mu}S_{21}(Q)\gamma^{\nu}S_{12}(K)])\tag{J}
\end{align}
\end{widetext}
In the limit of massless quarks, the longitudinal part of self energy is:
\begin{widetext}
\begin{align}
\Pi_{R}^{L}(P)=-iN_{f}g^{2}\int \frac{d^{4}K}{(2\pi)^{4}}(q_{0}k_{0}+\mathbf{q.k})\left [ \tilde{\Delta}_{F}(Q)\tilde{\Delta}_{R}(K)+\tilde{\Delta}_{A}(Q)\tilde{\Delta}_{F}(K)+\tilde{\Delta}_{A}(Q)\tilde{\Delta}_{A}(K)+\tilde{\Delta}_{R}(Q)\tilde{\Delta}_{R}(K) \right ]\tag{K}
\end{align}
\end{widetext}
In weak coupling limit, the external momentum is much lower than the internal momentum, so the R self energy in the HTL approximation simplifies into~\cite{12}:
\begin{align}
\Pi_{R}^{L}(P)=\frac{4\pi N_{f} g^{2}}{(2\pi)^{4}}\int kdk\int d\Omega f_{F}(\mathbf{k})\frac{1-(\mathbf{\hat{k}.\hat{p}})^{2}}{(\mathbf{\hat{k}.\hat{p}}+\frac{\mathbf{p_{0}}+i\epsilon }{\mathbf{p}})^{2}}\tag{L}
\end{align}
After elaborating the distribution function, in an weakly anisotropic ($\xi$ is not equal to 0) medium, the R quark self energy becomes:
\begin{align}
\Pi_{R}^{L}(P)=\frac{g^{2}}{(2\pi)^{2}}N_{f}\sum_{i=0}^{1}\int_{0}^{\infty }k\Phi_{(i)}(k)dk\int_{-1}^{1}\Psi_{(i)}(s)ds\tag{M}
\end{align}
with
\begin{align}
\Phi_{(0)}(k)=n_{F}(k)\tag{N}
\end{align}
\begin{align}
\Phi_{(1)}(k)=-\xi n_{F}^{2}(k)\frac{ke^{k/T}}{2T}\tag{O}
\end{align}
\begin{align}
\Psi _{(0)}(s)=\frac{1-s^{2}}{(s+\frac{p_{0}+i\epsilon}{p})^{2}}\tag{P}
\end{align}
and,
\begin{align}
\Psi _{(1)}(s)=cos^{2}\theta_{p}\frac{s^{2}(1-s^{2})}{(s+\frac{p_{0}+i\epsilon}{p})^{2}}+\frac{sin^{2}\theta_{p}}{2}\frac{(1-s^{2})^{2}}{(s+\frac{p_{0}+i\epsilon}{p})^{2}}\tag{Q}
\end{align}
Here, the angle $\theta_{p}$ is defined the angle between the $\mathbf{n}$ and $\mathbf{p}$ and s=$\mathbf{\hat{k}.\hat{p}}$, after spoil into anisotropic pieces, the anisotropic and $\xi=0$ (isotropic) terms becomes:
\begin{widetext}
\begin{align}
\Pi_{R(aniso)}^{L}(P)=N_{f}\frac{g^{2}T^{2}}{6}\left ( \frac{1}{6} +\frac{cos2\theta_{p}}{2}\right )+\Pi_{R(iso)}^{L}(P)\left (cos2\theta_{p}-\frac{p_{0}^{2}}{2p^{2}}(1+3cos2\theta_{p}) \right )\tag{R}
\end{align}
\end{widetext}
and 
\begin{align}
\Pi_{R(iso)}^{L}(P)=N_{f}\frac{g^{2}T^{2}}{6}\left ( \frac{p_{0}}{2p}ln\frac{p_{0}+p\pm i\epsilon}{p_{0}-p\pm i\epsilon}-1 \right )\tag{S}
\end{align}
Thus, the gluon self energy found the both imaginary and real part, which are accountable for Landau damping and Debye Screening, respectively where is usually obtained from the R and advanced self energy and later is obtained from the F self energy alone.
\end{document}